\documentstyle[11pt]{article}

\columnwidth\textwidth

\begin{document}

\title{Perturbative gauge invariance: the electroweak theory}

\author{Michael D\"utsch, G\"unter Scharf\\
Institut f\"ur Theoretische Physik der 
Universit\"at Z\"urich\\
Winterthurerstrasse 190, 8057 Z\"urich, Switzerland}
\date{}

\maketitle
\begin{abstract}
The concept of perturbative gauge invariance
formulated exclusively by means of asymptotic fields is generalized to
massive gauge fields. Applying it to the electroweak theory leads to a
complete fixing of couplings of scalar and ghost fields and of the
coupling to leptons, in agreement with the standard theory. 
The W/Z mass ratio is also determined, as well as the chiral character
of the fermions. We start directly with massive gauge fields and leptons  
and, nevertheless, obtain a theory which satisfies perturbative gauge 
invariance. 
\end{abstract}
\newpage


\def\d{\partial}\def\=d{\,{\buildrel\rm def\over =}\,}
\def\dh{\mathop{\vphantom{\odot}\hbox{$\partial$}}}
\def\dl{\dh^\leftrightarrow}
\def\sqr#1#2{{\vcenter{\vbox{\hrule height.#2pt\hbox{\vrule width.#2pt 
height#1pt \kern#1pt \vrule width.#2pt}\hrule height.#2pt}}}}
\def\w{\mathchoice\sqr45\sqr45\sqr{2.1}3\sqr{1.5}3\,} 
\def\eps{\varepsilon}
\def\oe{\overline{\rm e}}
\def\onu{\overline{\nu}}
\def\ds{\hbox{\rlap/$\partial$}}
\def\lra{\leftrightarrow}

\section{Introduction}
There exist various different notions of gauge invariance. On the
classical level one considers global and local gauge invariance. The
former is a symmetry in the usual sense with a symmetry group acting
linearly on the fields so that the Lagrangian is invariant. If the
symmetry transformations are space-time dependent one speaks of local
gauge transformations. Then derivatives of the fields transform
differently than the fields themselves. In order to maintain invariance of
the Lagrangian it is now necessary to substitute the ordinary
derivatives by covariant derivatives containing additional fields, the 
gauge fields
with the right transformation properties under the local gauge group. It
is a guiding principle in gauge theories to introduce new fields and 
couplings to correct a violation of gauge invariance.

It seems to be simple to take over these concepts to the quantum level
if one uses path integral quantization methods. But the devil is in the
details: the functional integral (which is not under control) must be
restricted to sections transversal to the gauge orbits by choosing a
gauge fixing \cite{22}. This gives rise to the Faddeev-Popov determinant which
can be attributed to new fields called ghost fields. The total
Lagrangian with the gauge fixing ($L_{GF}$) and Faddeev-Popov ($L_{FP}$)
terms added is no longer gauge-invariant, but it is invariant under a
modified symmetry transformation called BRST transformation which
involves also the ghost fields.  It rests on the important fact that the BRST
variations of $L_{GF}$ and $L_{FP}$ cancel each other. 

Considering massive gauge theories the difficulties appear already at the 
classical level. The mass term
$$m^2A^\mu A_\mu\eqno(1.0)$$
($A^\mu$ denotes the vector potential) of the free Lagrangian is not gauge 
invariant. The usual way out is the Higgs mechanism. One starts with the 
massless gauge fields, couples them to scalar fields $\Phi$, introduces 
a gauge 
invariant potential $V(\Phi)$ which has a form that gives rise to spontaneous
symmetry breaking. In this process the gauge bosons acquire a mass $m>0$.

The perturbative quantization of such a spontaneously broken gauge theory 
works beautifully \cite{11}: renormalizability by power counting\footnote{This
means that (taking possible cancellations not into account)
the number of undetermined parameters, which need to be fixed by 
normalization conditions (see below), does not increase by going over to
higher orders of the perturbation series.} and (exact)
BRST-invariance can be fulfilled simultaneously. In addition the S-matrix has 
been shown to be independent of the parameters in the gauge fixing term
$L_{GF}$ and to be unitary on the space of physical states. The main 
physical result of this procedure is the appearance of at least one new 
physical field: the scalar Higgs field(s) which describe(s) the Higgs 
boson(s). However, since the latter has/have not been experimentally
discovered yet, there have been many attempts to construct a
consistent, Higgs-free, nonabelian gauge theory. All failed: either they are 
not renormalizable by power counting or they violate physical unitarity (see
\cite{32} for a review).

In spite of this success the Higgs mechanism has the conceptual drawback 
that it relies on semiclassical arguments. However, the fundamental theory 
should be the quantum theory, because the microscopic world is quantum in 
nature. The quantum theory should tell the classical theory how it goes
and not the other way round. Hence a {\it pure quantum formulation of gauge 
invariance} (without reference to classical field theory)
is wanted. In the framework of perturbation theory a solution 
has been proposed recently for QED \cite{2} and massless non-abelian gauge 
theories \cite{3,7}. In this paper we apply this method to massive gauge 
theories.
A crucial difference to the Higgs mechanism is that we already start with
the {\it massive} (asymptotic free) fields, which (apart from the ghost 
fields) describe the observable (massive) particles.
Spontaneous symmetry breaking 
plays no role in our approach. Such a procedure is compatible with our  
formulation of gauge invariance, because the latter does not involve 
the mass term (1.0).

Our formulation of gauge invariance is adapted to {\it causal perturbation 
theory}, which goes back to St\"uckelberg, Bogoliubov and Shirkov, and 
Epstein and Glaser \cite{1} and offers considerable advantages. In 
this theory the S-matrix is constructed inductively order by order in the
form
$$S(h)=1+\sum_{n=1}^\infty{1\over n!}\int d^4x_1\ldots d^4x_n\, 
T_n(x_1,\ldots ,x_n)h(x_1)\ldots h(x_n),\eqno(1.1)$$
where $h(x)$ is a tempered test function that switches the interaction.
Hence the problems of the long-distance behaviour (infrared divergences) 
are absent. They appear in the adiabatic limit $h(x)\rightarrow 1$ (see
below).
The $T_n(x_1,\ldots ,x_n)$ are operator-valued distributions. The first order
expression $T_1(x)$ must be given to specify the model. It is a Wick polynomial
in the incoming free fields and corresponds to the interaction Lagrangian.
The higher orders $T_n,\>n\geq 2$ are constructed inductively, where the main
input is {\it causality}:
$$T_n(x_1,\ldots ,x_n)=T_l(x_{\pi 1},\ldots ,x_{\pi l})T_{n-l}(x_{\pi (l+1)},
\ldots ,x_{\pi n})$$
if $\{x_{\pi 1},\ldots ,x_{\pi l}\}\cap (\{x_{\pi (l+1)},\ldots ,x_{\pi n}\}
+\bar V^-)=\emptyset$ ($\pi$ is a permutation and $\bar V^-$ is the closed
backward light-cone). This
means that $T_n(x_1,\ldots ,x_n)$ is a
(well-defined) {\it time-ordered product} of 
$T_1(x_1),T_1(x_2),\ldots ,T_1(x_n)$:
$T_n(x_1,\ldots ,x_n)=T[T_1(x_1)T_1(x_2)\ldots T_1(x_n)]$.
The $T_n$'s are expanded in normally ordered products of the free incoming 
fields. Interacting fields do not appear. In the inductive step (from $(n-1)$
to $n$) $T_n$ is uniquely determined by causality up to the total diagonal
$x_1=x_2=\ldots =x_n$. The essential difficulty is the extension of the 
time-ordered products to the total diagonal. Epstein and Glaser do this in 
a hidden way by the splitting of a causal distribution \cite{1,2}. This
method has the practical advantage that the splitting can be done by
computing a dispersion integral in momentum space. However, the 
extension can also be done directly (in $x$-space) \cite{19,20,35}. The 
latter method is well suited for a generalization to curved space times.
If the extension is correctly done, ultraviolet divergences don't appear 
which is certainly a merit over conventional methods. In general the 
extension is non-unique: a local distribution (i.e. with support on the total 
diagonal) may be added. The form of this distribution is strongly restricted 
by renormalizability by power counting and by symmetry requirements, e.g.
Poincar\'e covariance. A big part of the present two papers deals with the 
restrictions coming from gauge invariance. The choice of a special extension 
to the total diagonal (or a special splitting solution) is called 
'normalization'. In general renormalizability by power counting
and the symmetry requirements
do not suffice to fix the theory, additional normalization
conditions are needed.

Regarding gauge theories in the causal approach it is necessary to 
formulate gauge invariance
in terms of the $T_n$'s because these are the fundamental objects.
For a pure massless Yang-Mills theory $T_1$ is given by
$$T_1=igf_{abc}(A_{\mu a}A_{\nu b}\d^\nu A_c^\mu 
-A_{\mu a}u_b\d^\mu\tilde u_c)\>,\eqno(1.2)$$
where $f_{abc}$ are (totally antisymmetric) structure constants of a 
non-abelian Lie algebra,
$A_{\mu a}(x)$ are the gauge potentials, and $u_a(x)$ 
and $\tilde u_a(x)$ are the fermionic ghost fields. {\it All products of
field operators throughout are normally ordered}. For notational simplicity 
we omit the double dots.\footnote{In section 3.1 
it is explained that (1.2) is essentially the only possibility for $T_1$
which is gauge invariant. The reader will probably miss the quadrilinear 
terms of the interaction Lagrangian. Since (1.1) is a perturbation series 
in the coupling constant $g$ and not in Planck's constant 
$\hbar$ (loop expansion)
these terms are of second order. They enter the present formalism in the 
normalization of the tree diagrams at second order. We shall see that 
they are uniquely fixed by gauge invariance.}
If all fields are massless, gauge invariance can be defined as follows.
Let
$$Q\=d \int d^3x\,(\d_{\nu}A_a^{\nu}{\dl}_0u_a)
\eqno(1.3)$$
be the generator of (free) gauge transformations \cite{3}, which was
first introduced by Kugo and Ojima \cite{14}. It implements the 
BRST-transformation $s$ of the asymptotic free fields: $s(\phi)=[Q,\phi]_\mp$, 
where we have the anticommutator if $\phi$ is a ghost field.
Then the commutator
$$[Q,T_1(x)]=gf_{abc}\d_\nu[A_{\mu a}u_b(\d^\nu A_c^\mu-\d^\mu A_c^\nu)
(x)+{1\over 2}u_au_b\d^\nu\tilde u_c(x)]$$
$$\=d i\d_\nu T^\nu_{1/1}(x)\eqno(1.4)$$
is a divergence. This is gauge invariance at first order, which is simply
the BRST condition on the coupling at order $g$ modulo
the free field equations. Note however that in
(1.4) the gauge variation of the ghost term is combined with the gauge
variation of the Yang-Mills term, in contrast to the usual combination
of $L_{FP}$ and $L_{GF}$ mentioned above. Furthermore, in (1.4) the wave
equations for the free fields have been used. All that is quite different from
the BRST approach \cite{11} which deals with the interacting fields.  We call
$T_{1/1}^\nu$ the $Q$-vertex. It does not describe a physical coupling,
but is a mathematical tool to formulate gauge invariance. Gauge
invariance at $n$-th order is now defined by 
$$d_QT_n\=d [Q,T_n]=i\sum_{l=1}^n{\d\over\d x_l^\nu}T^\nu_{n/l} 
(x_1,\ldots x_l\ldots x_n),\eqno(1.5)$$
where
$$T^\nu_{n/l}(x_1,\ldots x_l\ldots x_n)\equiv T[T_1(x_1)\ldots 
T^\nu_{1/1}(x_l)\ldots T_1(x_n)],$$
i.e. in $T_{n/l}^\nu$ the $l$-th vertex is a 
$Q$-vertex, all other $n-1$ vertices are ordinary Yang-Mills vertices (1.2).
The inductive construction of the time-ordered products is thus carried 
out for a bigger theory, involving also products with one $Q$-vertex.
Choosing proper normalizations in the construction of the
$T_n, T_{n/l}^\nu$, gauge invariance (1.5) can be fulfilled to all orders.
This has been proved in any $\lambda$-gauge (i.e. for any gauge fixing
term $L_{GF}={\lambda\over 2}\sum_a (\partial_\mu A^\mu_a)^2,\>\lambda >0$)
for massless nonabelian $SU(N)$ gauge theories \cite{3,7,24}.
By means of this result one can prove the $\lambda$-independence
(i.e. gauge independence) of the physical S-matrix up to divergence terms
\cite{24}. Such terms vanish in the adiabatic limit
$$S=\lim_{\epsilon\to 0} S(h_\epsilon),\quad\quad  h_\epsilon (x)
\equiv h(\epsilon x),\eqno(1.6)$$
provided the latter exists. However, this assumption
holds true only in pure massive
gauge theories (see below). But we emphasize that (1.5) is a {\it local}
condition which is independent of the adiabatic limit. If the Green's
functions exist, i.e. one can integrate out the inner vertices of the
diagrams in $T_n(x_1,\dots ,x_n)$ with $h(x)\equiv 1$, then (1.5) implies the 
usual Slavnov-Taylor identities \cite{25}. In other words the identity (1.5)
contains the full information of the usual formulation of non-abelian
gauge invariance. In addition (1.5) determines the possible structure of 
the model to a large extent. It is the main content of the present two 
papers to work this out for the electroweak theory.

An essential property of $Q$ is its nilpotency
$$Q^2=0\quad\Leftrightarrow \quad{\rm Ran}\,Q\subset{\rm Ker}\,Q,\eqno(1.7)$$
which reflects the same property of the BRST-transformation: $s^2=0$.
The quantization of the free gauge fields in a $\lambda$-gauge requires an
indefinite metric space. One can show that this indefinite inner product 
$<.,.>$ is
positive semidefinite on ${\rm Ker}\,Q$, and that the vectors in 
${\rm Ker}\,Q$ with "norm" zero are precisely the elements of ${\rm Ran}\,Q$
\cite{7,10}. Hence
$${\cal H}_{\rm phys}\equiv {{\rm Ker}\,Q\over {\rm Ran}\,Q},\quad\quad
<[\phi],[\psi]>\equiv <\phi,\psi>\eqno(1.8)$$
($[\phi]$ denotes the equivalence class of $\phi$) is a pre Hilbert space.
${\cal H}_{\rm phys}$ is interpreted as the space of physical states.
Provided the adiabatic limit exists, gauge invariance (1.5) implies
$[Q,S]=0$, because the divergence terms on the r.h.s. in (1.5) vanish in this
limit. In addition one can choose the normalizations such that $S^*S=
{\bf 1}=SS^*$ ($*$ means the adjoint with respect to the indefinite inner
product). Then, $S$ induces a well-defined operator $[S]$ on the factor
space ${\cal H}_{\rm phys}$ by
$$[S][\phi]\equiv[S\phi],\eqno(1.9)$$
which satisfies physical unitarity: $[S]^*[S]={\bf 1}=[S][S]^*$ \cite{26}.

The drawback of the present S-matrix formalism is that the physical 
interpretation is only clear if the adiabatic limit (1.6) exists. In
massless nonabelian gauge theories this is certainly not the case; a 
{\it local} construction of the observables (which avoids the adiabatic limit)
is closer to the physical reality. Such a construction is in preparation
\cite{30}. This construction relies on the validity of gauge invariance 
in the sense of (1.5), what shows the usefulness of (1.5).

If all fields are massive the adiabatic limit of the S-matrix (1.6) exists 
in the strong operator topology (provided a correct mass and wave function 
normalization is done) \cite{28} and, hence, the present formalism 
has a direct physical interpretation. This fact and the preprint version
of the present two papers (as well as a third paper of the authors \cite{6})
has inspired Schroer \cite{27}, Grigore \cite{23}\footnote{However, 
there is a mistake in Grigores paper which can be described as follows
(by using the terminology introduced below): the Higgs field(s) 
is/are treated as scalar partner(s) (with
arbitrary mass $m_{H}\geq 0$) of the massless gauge field(s), which does/do
not appear in $Q$ and, hence, is/are physical. By chance this works for the
electroweak theory (there is one massless gauge field and one Higgs field is
needed). But e.g. in the case of pure massive gauge theories, there
would be no Higgs field and such a model violates gauge invariance (1.5)
to second order.} and one of the authors and
Schroer \cite{26} to take the following point of view: they do not require 
gauge invariance (1.5) (as a physical principle), instead they only require 
physical consistency. The latter means that the S-matrix induces a 
well-defined (unitary) operator on ${\cal H}_{\rm phys}$ by (1.9). This is
equivalent to 
$$[Q,S]P_K=0,\eqno(1.10)$$
where $P_K$ is the projector on ${\rm Ker\,Q}$. Then they show that (1.9)
determines the possible structure of the model to the same extent as
gauge invariance (1.5). They do this by repeating (or refering to)
the calculations in the present two papers. In \cite{26} the requirement
(1.9) is replaced by the existence of certain observables, i.e. interacting 
fields of a certain form which commute with $Q$. According to these references
the (physical) Higgs field is needed for physical consistency. 
\footnote{This point of view is related to the (old) strategy to derive the 
electroweak theory from unitarity (more precisely from the bounds on the 
high-energy behavior of cross sections which are due to unitarity).
In this way unitarity was first an argument against the current-current
interaction of Fermi. Later on, it motivated the introduction of the
Z-boson and the corresponding neutral current, and also of the Higgs
boson \cite{15,16,17}. In the present two papers unitarity is not used for the
construction of the theory, but can be proven if the theory is gauge
invariant (1.5) with a nilpotent $Q$ (1.7).}

In the electroweak theory we do not expect that the strong adiabatic limit
(1.6) exists, because of the vanishing photon mass\footnote{Due to this fact 
there seem to be a loophole in the derivation 
of the standard model in \cite{23}.}.
In QED (which is part of the electroweak theory) the adiabatic limit of 
Green's functions exists \cite{31}, but (to our knowledge) the strong limit
of the S-matrix (1.6) does not. A further argument against the adiabatic 
limit is the existence of unstable physical particles ($W$-, $Z$-boson, 
$\mu$- and $\tau$-leptons). To describe decay and scattering processes 
involving such particles we work with a $h(x)$ which has compact support
in time.

In working out the formulation (1.5) of quantum gauge invariance
for massive gauge fields the question arises whether this identity
must be broken (by mass terms) or can be maintained by introducing new 
fields and couplings.
In the first case the gauge principle is not strong enough
to determine the whole theory, some additional symmetry-breaking
mechanism must be added. In the second case the gauge principle is
strong and fixes the couplings. We are going to show that the second
alternative can indeed be realized in the causal approach.
The only possibility we know to maintain gauge invariance (1.5) 
in the massive theory with a nilpotent $Q$ \footnote{If one gives up
the nilpotency of $Q$ (1.7) perturbative gauge invariance (1.5) can be
satified without introducing any bosonic scalar fields \cite{29}.
Such an approach is similar to the Curci-Ferrari model \cite{33}.
But these models violate physical unitarity due to $Q^2\not= 0$.}
is to introduce at least one additional physical field (the scalar Higgs
field), with vanishing
vacuum expectation values. So we directly arrive at the final form
of the electroweak theory with massive gauge fields and leptons
which, nevertheless, is manifestly gauge invariant after construction.
In an earlier paper \cite{6} we have demonstrated our 
method in the simple case of the abelian Higgs model. Here we treat 
the full electroweak theory in all details and we show how this big
theory comes out as a consequence of perturbative gauge invariance.

It is our aim to find a theory with massive fields $A_a, u_a, \tilde
u_a$ without violating the properties (1.4-5)
and (1.7). If we simply substitute
the massless fields in the above expressions by massive ones, then all
three relations get lost. To restore (1.7) we introduce scalar fields
$\Phi_a$ and modify the expression (1.3) of $Q$. This is discussed in
the following section. Then, to get gauge invariance to first order
(1.4) we have to couple these (unphysical) scalar fields to the $A$'s 
and $u$'s in a suitable way. These couplings are not of Yang-Mills
type, that means they are not proportional to $f_{abc}$,
due to the non-equality of the masses of the gauge bosons. 
We will find them by making a general ansatz and using gauge
invariance to determine the parameters. In Sect.3 this is done at first
order which completely fixes the couplings of the unphysical scalars.
However, gauge invariance at second order requires the introduction of
(at least one) additional physical scalar, the 'Higgs' field. 
In Sect.4 its coupling
is derived and the usual results for the boson masses are obtained. 
In Sect.5 we discuss the coupling to leptons. The chiral character
of these couplings is not put in, but comes out as a consequence of
gauge invariance. All couplings agree precisely with what is obtained 
from the Glashow-Salam-Weinberg theory \cite{8,9} 
{\it after symmetry breaking}. Spontaneous symmetry breaking plays no 
role in our theory, since we directly start with massive asymptotic
gauge fields and leptons. We do not consider gauge invariance
at higher orders ($n>2$) here. In part II of this paper we shall
discuss third order gauge invariance which enables us to derive the
Higgs potential.

Our procedure has some similarity with the 'first order formalism'
of Deser \cite{34}, which relies on the classical Noether method.
Similar to our argumentation, where the free fields and the gauge
charge $Q$ (1.3) are the main input, this formalism starts from the gauge
transformations of the free fields and derives by 'consistency' the full
Lagrangian of the interacting theory. However, this is pure classical 
Lagrangian field theory, whereas we argue completely on the quantum level.

\vskip 1cm
\section{Free Theory and Infinitesimal Gauge Transformations}
\vskip 1cm
A first step towards a gauge theory with massive gauge fields in the
causal framework was recently made by F.Krahe \cite{10}. In this section we
essentially follow his arguments.  Unfortunately, his final theory  
was not gauge invariant at second order because he left out the  
physical scalar fields.

The choice of a gauge enters our formalism in the expression for the
commutator (or propagator) of the asymptotic free gauge fields.
\footnote{In massless gauge theories a gauge fixing is necessary because 
otherwise the propagator does not exist. In the massive case the situation 
is different: without a gauge fixing term $L_{GF}$ we have the 
Proca field. The corresponding propagator exists:
$$\hat D^F_m(k)\sim (g^{\mu\nu}-{k^\mu k^\nu\over m^2}){1\over k^2-m^2
+i\epsilon}$$ 
for the Feynman propagator in momentum space. But the
${k^\mu k^\nu\over m^2}$-term gives rise to a bad ultraviolet behavior,
which destroys renormalizability by power counting in nonabelian
theories.} We choose
the Feynman gauge $\lambda =1$ (where $L_{GF}={\lambda\over 2}\sum_a 
(\partial_\mu A^\mu_a)^2$), in which the free field equation is the 
Klein-Gordon equation and the commutator/propagator has a very simple form:
$$(\w +m_a^2)A_a^\mu(x)=0,\>[A_a^\mu(x),A_b^\nu(y)]_-=i\delta_{ab}g^
{\mu\nu}D_{m_a}(x-y).\eqno(2.1)$$
($D_m$ is the Jordan-Pauli distribution to the mass $m$.)
The ghost field $u_a$ must have the same mass as the corresponding gauge
field $A^\mu_a$. Otherwise the free BRST-current $j_\mu\equiv\d_{\nu}A_a^{\nu}
{\dl}_\mu u_a$ would not be conserved and the charge $Q=\int d^3x\,j_0$
(1.3) neither. Thus
$$(\w +m_a^2)u_a(x)=0=(\w +m_a^2)\tilde u_a(x)\eqno(2.2)$$
$$\{u_a(x),\tilde u_b(y)\}_+=-i\delta_{ab}D_{m_a}(x-y),\eqno(2.3)$$
all other commutators vanish. Due to the ${1\over k^2}$-behavior of the 
propagators for $k^2\rightarrow\infty$ (in momentum space) the theory is 
renormalizable by power counting if the mass dimension of $T_1$ is $\leq 4$.
The prove is the same as in the massless case (\cite{3} (1994), sect.2.2).
Calculating the square of $Q$ (1.3) by means of the anticommutator we find
$$Q^2={1\over 2}\{Q,Q\}={1\over 2}\int d^3x\int d^3y [\d_\mu A_a^\mu
(x),\d_\nu A_b^\nu (y)]{\dl}_{x_0}{\dl}_{y_0} u_a(x)u_b(y)
\eqno(2.4)$$
which is different from 0, because of $[\d_\mu A^\mu_a(x),\d_\nu
A^\nu_b(y)] = i\delta_{ab}m_a^2D_{m_a}(x-y)$. To restore the nilpotency we add to every massive
gauge vector field $A_a^\mu(x)$ a scalar partner $\Phi_a(x)$ with the
same mass. If the gauge field is massless it needs no partner.
The scalar fields are quantized according to
$$(\w +m_a^2)\Phi_a(x)=0,\>[\Phi_a(x),\Phi_b(y)]=-i\delta_{ab}
D_{m_a}(x-y).\eqno(2.5)$$
Due to the observation
$$[\d_\mu A^\mu_a(x)+m_a\Phi_a(x),\d_\nu A^\nu_b(y)+m_b\Phi_b(y)]=0$$
we get a nilpotent $Q$ by replacing $\d_\nu A^\nu_a$ by $(\d_\nu
A^\nu_a+m_a\Phi_a)$ in (1.3) \footnote{The equality of the masses of the 
gauge field $A_a$ and the partners $u_a,\,\tilde u_a$ and $\Phi_a$ is a 
speciality of the Feynman gauge. In an arbitrary $\lambda$-gauge the 
mass-square of 
$u_a,\,\tilde u_a$ and $\Phi_a$ is ${m_a^2\over\lambda}$ and the nilpotent 
charge $Q$ is obtained by replacing $\d_\nu A^\nu_a$ by $(\lambda\d_\nu
A^\nu_a+m_a\Phi_a)$ in (1.3).}
$$Q\=d \int d^3x\,(\d_{\nu}A_a^{\nu}+m_a\Phi_a){\dl}_0u_a,\quad Q^2=0.
\eqno(2.6)$$

The gauge charge $Q$ defines a gauge variation according to
$$d_Q F\=d QF-(-1)^{n_F}FQ,$$
where $n_F$ is the number of ghost fields in the Wick monomial $F$.
With the modified gauge charge (2.6) we get the following gauge variations of
the fundamental fields
$$d_QA_a^\mu(x)=i\d^\mu u_a(x),\quad d_Q\Phi_a(x)=im_a
u_a(x)\eqno(2.7)$$
$$d_Qu_a(x)=0,\quad d_Q\tilde u_a(x)=-i(\d_\mu A_a^\mu(x)+m_a\Phi_a(x)).
\eqno(2.8)$$
These infinitesimal gauge transformations are the BRST transformations of
the asymptotic free fields \cite{14,11}, but note the following
differences. The BRST transformations are defined for interacting
fields, whereas we work with asymptotic free fields only and establish
gauge invariance order by order. 
BRST invariance only holds if the quadratic free Lagrangian, the
gauge fixing term and the quartic term are also transformed. We have no 
such terms in $T_1$
(1.2) so that the compensations of terms in the gauge variations are
totally different.
If $m_a=0$ there is no scalar contribution in (2.8). The scalar fields
so far introduced are unhysical because they do not commute with $Q$ and,
therefore, their excitations do not
belong to the space of physical states ${\cal H}_{\rm phys}$ (1.8). 
But gauge invariance will force us to introduce an additional scalar field
$\Phi_0$ with arbitrary mass $m_H$, 
$$(\w +m_H^2)\Phi_0(x)=0,\>[\Phi_0(x),\Phi_0(y)]=-iD_{m_H}(x-y),\eqno(2.9)$$
which does not occur in $Q$ and, hence, is
physical and its gauge variation vanishes.

In the electroweak theory we have the following fundamental fields: the
massless photon $A^\mu(x)$, the two W-bosons $W_1^\mu(x), W_2^\mu(x)$ and the
Z-boson $Z^\mu(x)$. Note that in the causal theory the S-matrix is
expressed by the true outgoing asymptotic fields with definite masses 
which can directly be
used to generate physical scattering states. Consequently, we have to
work with the gauge fields after rotation by the electroweak mixing angle
$\Theta$. The three massive gauge fields have three unphysical scalar
partners $\Phi_1, \Phi_2, \Phi_3$. In addition we need one physical
scalar $\Phi_0$. We group the fields according to the color index $a$:
$$\vbox{
\halign{#\quad\quad&#\quad&#\quad&#\quad&#\quad&#\quad\cr 
a=&0,&1,&2,&3\cr $A_a^\mu $& $A^\mu$& $W_1^\mu$&
$W_2^\mu$& $Z^\mu$\cr $m_a$& 0& $m_1$& $m_2$& $m_Z$& $m_H$\cr $u_a$&$u_0$ 
&$u_1$&$u_2$&$u_3$\cr $\tilde u_a$&$\tilde u_0$ 
&$\tilde u_1$&$\tilde u_2$&$\tilde u_3$\cr
$\Phi_a$&$ $&$\Phi_1$&$\Phi_2$&$\Phi_3$&$\Phi_0$\cr}}
\eqno(2.10)$$
The 'Higgs' field $\Phi_0$ is indispensable for gauge invariance: While gauge
invariance at first order can be achieved without $\Phi_0$, this is
impossible at second order; here one needs (at least) one physical scalar 
field. For simplicity we only consider one Higgs field. To reduce the
number of cases in the following discussion we include the Higgs field
$\Phi_0$ in the family with color index $a=0$, although it is not the
scalar partner of the photon (cf. footnote 3). 

According to (2.6-8) the fields have the following gauge variations
$$d_QA^\mu =i\d^\mu u_0,\> d_Q W_{1,2}^\mu=i\d^\mu u_{1,2},\>
d_Q Z^\mu=i\d^\mu u_3\eqno(2.11)$$
$$d_Q\Phi_0=0,\>d_Q\Phi_{1,2}=im_{1,2}u_{1,2},\>d_Q\Phi_3=im_Zu_3
\eqno(2.12)$$
$$d_Qu_a=0,\quad a=0,1,2,3$$
$$d_Q\tilde u_0=-i\d_\mu A^\mu,\>d_Q\tilde u_{1,2}=-i(\d_\mu W_{1,2}
^\mu+m_{1,2}\Phi_{1,2})$$
$$d_Q\tilde u_3=-i(\d_\mu Z^\mu+m_Z\Phi_3).\eqno(2.13)$$
The masses of the W- and Z-bosons are still arbitrary. The usual
relations
$$m_1=m_2=m_W,\quad m_W=m_Z\cos\Theta\eqno(2.14)$$
will come out later as a consequence of gauge invariance.

\section{Gauge Invariance at First Order}
\subsection{Yang-Mills Part}
We start from the usual Yang-Mills coupling (1.2). It has been shown 
\cite{18, 21}
that this is the most general coupling consistent with perturbative
gauge invariance (1.4), up to divergence couplings $\sim \d_\mu (A^\mu A_\nu
A^\nu)$, $\sim \d_\mu (A^\mu u\tilde u)$ and coboundaries $\sim
d_Q(u\tilde u\tilde u)$, $d_Q(\tilde u A_\mu A^\mu)$. Such terms do not
violate gauge invariance also in higher orders \cite{12}. Note that 
couplings of the fermionic ghost fields $u,\,\tilde u$ are needed to 
satisfy (1.4).
The $f_{abc}$ in (1.2) must be totally antisymmetric. Gauge invariance
(1.5) for second order tree diagrams requires that they satisfy the
Jacobi identity. Hence, they are structure constants of a Lie algebra.
Quartic couplings in $T_1$ are not gauge invariant \cite{23} (cf. footnote 2).

In the electroweak theory the
structure constants $f_{abc}$ correspond to a
rotated basis of the $SU(2)\times U(1)$ Lie algebra and are equal to
$$f_{210}=\sin\Theta,\quad f_{321}=\cos\Theta,\quad f_{310}=0, 
\quad f_{320}=0,\eqno(3.1)$$
all other constants follow by antisymmetry. Such a rotation is of
physical significance in the massive case because we have assigned
definite masses to the basic fields.
The Yang-Mills part (1.2) remains unchanged in the massive theories. We
write the two terms of (1.2) as $T_1^A+T_1^u$ and calculate their gauge
variations. The transformation of the result to a divergence form is
exactly the same as in the zero mass case \cite{3}. The operator $d_Q$
operates on the algebra of normally ordered products of free fields
(including the Wick monomials) as a graded derivation, i.e. the product 
rule reads:
$$d_Q(FG)=(d_QF)G+(-1)^{n_F}Fd_QG,$$
where $n_F$ is the number of ghost fields $u, \tilde u$ in $F$. Then we get
$$d_QT_1^A=-gf_{abc}\Bigl(\d_\mu u_aA_{\nu b}\d^\nu A_c^\mu$$
$$+A_{\mu a}\d_\nu u_b\d^\nu A_c^\mu+A_{\mu a}A_{\nu b}\d^\nu\d^\mu u_c\Bigl).
\eqno(3.2)$$
The last term vanishes due to antisymmetry. To transform the expression
to divergence form we always take out the derivative of the ghost
fields:
$$d_QT_1^A=-gf_{abc}\Bigl(\d_\mu(u_aA_{\nu b}\d^\nu A_c^\mu)-u_a\d_\mu
A_{\nu b}\d^\nu A_c^\mu-u_aA_{\nu b}\d_\mu\d^\nu A_c^\mu$$
$$+\d_\nu(A_{\mu a}u_b\d^\nu A_c^\mu)-\d_\nu A_{\mu a}u_b\d^\nu A_c^\mu
-A_{\mu a}u_b\w A_c^\mu\Bigl).$$
Again, the second and fifth term vanishes by antisymmetry, in the last
one we use the field equation
$$d_QT_1^A=-gf_{abc}\Bigl(\d_\nu(u_aA_{\mu b}\d^\mu A_c^\nu)+\d_\nu(A_{\mu a}
u_b\d^\nu A_c^\mu)\eqno(3.3a)$$
$$+A_{\mu a}u_b\d^\mu\d_\nu A_c^\nu+m_c^2A_{\mu a}u_bA_c^\mu\Bigl).
\eqno(3.3b)$$

For the variation of the ghost term in (1.2) we get
$$d_QT_1^u=-gf_{abc}\Bigl(-\d_\mu u_au_b\d^\mu\tilde u_c-A_{\mu a}u_b
\d^\mu\d_\nu A_c^\nu$$
$$-m_cA_{\mu a}u_b\d^\mu\Phi_c\Bigl).\eqno(3.4)$$
Here the second term cancels the first in (3.3b). In the first term, say
$T_{11}^u$, we again take out the derivative from the ghost field
$$T^u_{11}=-gf_{abc}\Bigl(-\d_\mu (u_au_b\d^\mu\tilde u_c)+u_a\d_\mu
u_b\d^\mu\tilde u_c$$
$$+u_au_b\w\tilde u_c\Bigl).\eqno(3.5)$$
The second term is the negative of the left-hand side. Using the field
equation in the last term we find
$$T^u_{11}={g\over 2}f_{abc}\Bigl(\d_\nu (u_au_b\d^\nu\tilde u_c)
+m_c^2u_au_b\tilde u_c\Bigl).\eqno(3.6)$$
Summing up we have obtained a divergence apart from the mass terms:
$$d_Q(T_1^A+T_1^u)=gf_{abc}\biggl[\d_\nu\Bigl(A_{\mu a}u_b(\d^\mu A_c^\nu
-\d^\nu A_c^\mu)+{1\over 2} u_au_b\d^\nu\tilde u_c\Bigl)$$
$$-m_c^2A_{\mu a}u_bA_c^\mu+{1\over 2} m_c^2u_au_b\tilde u_c+m_cA_{\mu a}u_b
\d^\mu\Phi_c\biggl].\eqno(3.7)$$
Of course, this agrees with (1.4) if all masses are put =0. The mass
terms violate gauge invariance. To compensate them couplings of the 
scalar fields must be introduced. 

For the following we write down (3.7) explicitly for the electroweak
theory:
$$d_Q(T_1^A+T_1^u)=-g\d_\mu \biggl[\sin\Theta u_1 W_2^\nu\d^\mu A_\nu 
+\cos\Theta u_2Z^\nu\d^\mu W_{1\nu}\eqno(3.7.1)$$
$$-\cos\Theta u_3W_2^\nu\d^\mu W_{1\nu}-\sin\Theta u_0W_2^\nu\d^\mu W_{1\nu}
-\sin\Theta u_2 W_1^\nu\d^\mu A_\nu\eqno(3.7.2)$$
$$+\sin\Theta u_2A^\nu\d^\mu W_{1\nu}+\cos\Theta u_3W_1^\nu\d^\mu W_{2\nu}
-\cos\Theta u_1Z^\nu\d^\mu W_{2\nu}\eqno(3.7.3)$$
$$+\sin\Theta u_0W_1^\nu\d^\mu W_{2\nu}-\sin\Theta u_1A^\nu\d^\mu W_{2\nu}
+\cos\Theta u_1W_2^\nu\d^\mu Z_\nu
-\cos\Theta u_2W_1^\nu\d^\mu Z_\nu\eqno(3.7.4)$$ 
$$+\sin\Theta u_2 W_1^\nu\d_\nu A^\mu-\sin\Theta u_1W_2^\nu\d_\nu A^\mu
+\cos\Theta u_3W_2^\nu\d_\nu W_{1\mu}$$
$$-\cos\Theta u_2Z^\nu\d_\nu W_{1\mu}-\sin\Theta u_2A^\nu\d_\nu W_{1\mu}
+\sin\Theta u_0W_2^\nu\d_\nu W_{1\mu}$$
$$+\cos\Theta u_1Z^\nu\d_\nu W_{2\mu}-\cos\Theta u_3W_1^\nu\d_\nu W_{2\mu}
+\sin\Theta u_1A^\nu\d_\nu W_{2\mu}$$
$$-\sin\Theta u_0W_1^\nu\d_\nu W_{2\mu}+\cos\Theta u_2W_1^\nu\d_\nu Z^\mu
-\cos\Theta u_1W_2^\nu\d_\nu Z^\mu\eqno(3.7.5)$$
$$+\sin\Theta u_1u_2\d^\mu\tilde u_0+\cos\Theta u_2u_3\d^\mu\tilde u_1
-\sin\Theta u_0u_2\d^\mu\tilde u_1$$
$$+\cos\Theta u_3u_1\d^\mu\tilde u_2+\sin\Theta u_0u_1\d^\mu\tilde u_2
+\cos\Theta u_1u_2\d^\mu\tilde u_3\biggl]\eqno(3.7.6)$$
$$-g\biggl[\cos\Theta(m_1^2-m_Z^2)u_2W_1^\nu Z_\nu-\cos\Theta(m_2^2-m_Z^2)
u_1W_2^\nu Z_\nu$$ 
$$+\sin\Theta(m_1^2u_2W_1^\nu-m_2^2u_1W_2^\nu)A_\nu\eqno(3.7.7)$$
$$+\cos\Theta(m_2^2-m_1^2)u_3W_{1\nu}W_2^\nu+\sin\Theta(m_1^2-m_2^2)u_0
W_{1\nu}W_2^\nu\eqno(3.7.8)$$
$$+\cos\Theta m_1(u_3W_2^\nu-u_2Z^\nu)\d_\nu\Phi_2-\sin\Theta m_1(u_2A^\nu-u_0
W_2^\nu)\d_\nu\Phi_1$$ 
$$+\cos\Theta m_2(u_1Z^\nu-u_3W_1^\nu)\d_\nu\Phi_2\eqno(3.7.9)
$$
$$+\sin\Theta m_2(u_1A^\nu-u_0W_1^\nu)\d_\nu\Phi_2+\cos\Theta m_Z(u_2W_1^\nu
-u_1W_2^\nu)\d_\nu\Phi_3\eqno(3.7.10)$$ 
$$-\cos\Theta m_1^2\tilde u_1u_3u_2-\sin\Theta m_1^2\tilde u_1u_0u_2+\cos\Theta
m_2^2\tilde u_2u_3u_1\eqno(3.7.11)$$
$$+\sin\Theta m_2^2\tilde u_2u_0u_1+\cos\Theta m_Z^2\tilde u_3u_1u_2\biggl]
\eqno(3.7.12)$$
The first square bracket (3.7.1-6) defines the Yang-Mills part 
$T_{1/1}^{A\mu}+T_{1/1}^{u\mu}$ of the $Q$-vertex $T_{1/1}^\mu$.
\vskip 0.5cm
\subsection{Scalar Couplings in Sector (1,2,3)}
\vskip 0.5cm
We now introduce couplings to the scalar fields in order to achieve
gauge invariance. We require that they are Lorentz scalars, have ghost
number zero and are trilinear in the fields. Since the gauge variation 
$d_Q$ (2.7-8) does not mix fields
with different $a=0,1,2,3$, we can solve the problem step by step in
small sectors. Let us consider all coupling terms containing exactly one
field with $a=1$, one field with $a=2$ and the third one with $a=3$.
They form the sector (1,2,3). Since the mass terms (3.7.7-12) which must
be compensated are not of Yang-Mills type, we make the following ansatz
for the scalar coupling in this sector:
$$T_1^{123}=ig\Bigl[a_1W_1^\mu(\Phi_2\d_\mu\Phi_3-\Phi_3\d_\mu\Phi_2)+
a_2W_2^\mu(\Phi_3\d_\mu\Phi_1-\Phi_1\d_\mu\Phi_3)$$
$$+a_3Z^\mu(\Phi_1\d_\mu\Phi_2-\Phi_2\d_\mu\Phi_1)+b_1W_{1\mu}W_2^\mu\Phi_3
+b_2W_{2\mu}Z^\mu\Phi_1$$
$$+b_3Z^\mu W_{1\mu}\Phi_2+c_1\tilde u_1u_2\Phi_3+c_1'\tilde u_2u_1\Phi_3
+(c_2\tilde u_2u_3+c_2'\tilde u_3u_2)\Phi_1$$
$$+(c_3\tilde u_3u_1+c_3'\tilde u_1u_3)\Phi_2+d_1\Phi_1\Phi_2\Phi_3
\Bigl].\eqno(3.8)$$
A general ansatz contains additional terms $\sim\d_\nu A_a^\nu
\Phi_b\Phi_c$ and $\sim A^\nu_a(\d_\nu\Phi_b\Phi_c+\Phi_b\d_\nu\Phi_c)$.
It can be reduced to (3.8) by adding divergences $\d_\nu (A_a^\nu
\Phi_b\Phi_c)$ and coboundaries $d_Q(\tilde u_a\Phi_b\Phi_c)$.
Such addition does not change gauge invariance (at least at low orders) 
\cite{12}.
The form (3.8) has the nice property that $d_Q T_1$ can be transformed
to divergence form by simply taking out the derivatives of the ghost
fields as described above.

As before, we calculate $d_QT_1^{123}$ and form a divergence by taking
out the derivatives of the ghost fields. The result is
$$d_QT_1^{123}=-g\d^\mu\Bigl[a_1u_1(\Phi_2\d_\mu\Phi_3-\Phi_3\d_\mu
\Phi_2)$$
$$+a_1m_Zu_3W_1^\mu\Phi_2-a_1m_2u_2W_1^\mu\Phi_3+a_2u_2(\Phi_3\d_\mu
\Phi_1-\Phi_1\d_\mu\Phi_3)+a_2m_1u_1W_{2\mu}\Phi_3$$
$$-a_2m_Zu_3W_{2\mu}\Phi_1+a_3u_3(\Phi_1\d_\mu\Phi_2-\Phi_2 
\d_\mu\Phi_1)+a_3m_2u_2Z_\mu\Phi_1$$
$$-a_3m_1u_1Z_\mu\Phi_2+b_1u_1W_{2\mu}\Phi_3+b_1u_2W_{1\mu}\Phi_3
+b_2u_2Z_\mu\Phi_1$$
$$+b_2u_3W_{2\mu}\Phi_1+b_3u_3W_{1\mu}\Phi_2+b_3u_1Z_\mu\Phi_2\Bigl]
+T_{11}^{123}.\eqno(3.9)$$
The terms $T_{11}^{123}$ have no derivative on the ghost fields. If
combined with the corresponding terms in (3.7.7-3.7.12) their
coefficients must vanish, in order to have gauge invariance, because
these monomials are linear independent modulo divergences.

This leads to the following conditions
$$0=u_3W_{1\mu}W_2^\mu(\cos\Theta(m_2^2-m_1^2)+b_1m_Z)=u_1W_{2\mu} 
Z^\mu(\cos\Theta(m_Z^2-m_2^2)+b_2m_1)=$$ $$=u_2Z_\mu W_1^\mu(\cos\Theta
(m_1^2-m_Z^2)+b_3m_2)=
u_3W_1^\mu\d_\mu\Phi_2(-m_2\cos\Theta-2a_1m_Z-b_3)=$$ $$=u_2W_1^\mu
\d_\mu\Phi_3(m_Z\cos\Theta+2a_1m_2-b_1)=u_3W_2^\mu\d_\mu\Phi_1(m_1
\cos\Theta+2a_2m_Z-b_2)=$$
$$=u_1W_2^\mu\d_\mu\Phi_3(-m_Z\cos\Theta-2a_2m_1-b_1)=u_1Z^\mu\d_\mu
\Phi_2(m_2\cos\Theta+2a_3m_1-b_3)$$ $$=u_2Z^\mu\d_\mu\Phi_1(-m_1\cos 
\Theta-2a_3m_2-b_2)=
u_2\d_\mu W_1^\mu\Phi_3(a_1m_2-b_1-c_1)$$ $$=u_1\d_\mu W_2^\mu\Phi_3 
(-a_2m_1-b_1-c'_1)=u_3\d_\mu W_2^\mu\Phi_1(a_2m_Z-b_2-c_2)=$$
$$=u_2\d_\mu Z^\mu\Phi_1(-a_3m_2-b_2-c'_2)=u_1\d_\mu Z^\mu\Phi_2 
(a_3m_1-b_3-c_3)$$ 
$$=u_3\d_\mu W_1^\mu\Phi_2(-a_1m_Z-b_3-c'_3)=
u_1\Phi_2\Phi_3(a_1(m_Z^2-m_2^2)-c'_1m_2-c_3m_Z+d_1m_1)$$
$$=u_3\Phi_1\Phi_2(a_3(m_2^2-m_1^2)-c_2m_2-c'_3m_1+d_1m_3)
=\tilde u_1u_2u_3(m_1^2\cos\Theta+c_1m_Z-c'_3m_2)$$ 
$$=\tilde u_2u_3u_1(m_2^2\cos\Theta-c'_1m_Z+c_2m_1)
=\tilde u_3u_1u_2(m_Z^2\cos\Theta-c'_2m_1+c_3m_2)$$  
$$=u_2\Phi_1\Phi_3(a_2(m_1^2-m_Z^2)-c'_2m_Z-c_1m_1+d_1m_2)=0.\eqno(3.10)$$     
The first 15 equations imply
$$b_1={m_1^2-m_2^2\over m_Z}\cos\Theta,\>b_2={m_2^2-m_Z^2\over m_1} 
\cos\Theta,\>b_3={m_Z^2-m_1^2\over m_2}\cos\Theta,\eqno(3.11)$$   
$$a_1={m_1^2-m_2^2-m_Z^2\over 2m_2m_Z}\cos\Theta,\>   
a_2={m_2^2-m_Z^2-m_1^2\over 2m_1m_Z}\cos\Theta,\>   
a_3={m_Z^2-m_1^2-m_2^2\over 2m_1m_2}\cos\Theta,\eqno(3.12)$$
$$c_1={m_2^2-m_1^2-m_Z^2\over 2m_Z}\cos\Theta,\>      
c'_1={m_Z^2-m_1^2+m_2^2\over 2m_Z}\cos\Theta,\>
c_2={m_Z^2-m_2^2-m_1^2\over 2m_1}\cos\Theta,$$
$$c'_2={m_Z^2-m_2^2+m_1^2\over 2m_1}\cos\Theta,\>
c_3={m_1^2-m_2^2-m_Z^2\over 2m_2}\cos\Theta,\>
c'_3={m_1^2+m_2^2-m_Z^2\over 2m_2}\cos\Theta.\eqno(3.13)$$
The remaining relations are then automatically satisfied with arbitrary
masses and $d_1=0$.
\vskip 0.5cm
\subsection{Sectors (0,1,2), (1,1,2), (1,1,3), (1,3,3), etc.}
\vskip 0.5cm
Similar to (3.8) the ansatz for the coupling in the sector
(0,1,2) reads as follows
$$T_1^{012}=ig\Bigl\{a_4 A^\mu(\Phi_1\d_\mu\Phi_2-\Phi_2\d_\mu
\Phi_1)+a_5W_1^\mu(\Phi_2\d_\mu\Phi_0-\Phi_0\d_\mu\Phi_2)$$
$$+a_6W_2^\mu(\Phi_0\d_\mu\Phi_1-\Phi_1\d_\mu\Phi_0)+b_4A_\mu
W_1^\mu\Phi_2+b_5W_{1\mu}W_2^\mu\Phi_0$$
$$+b_6W_{2\mu}A^\mu\Phi_1+c_4\tilde u_0u_1\Phi_2+c_4'\tilde u_1u_0 
\Phi_2+c_5\tilde u_1u_2\Phi_0$$ 
$$+c_5'\tilde u_2u_1\Phi_0+c_6\tilde u_2u_0\Phi_1+c'_6\tilde
u_0u_2\Phi_1+d_2\Phi_0\Phi_1\Phi_2 \Bigl\}.\eqno(3.14)$$
Proceeding in the same way as in the last section we get for $b_4$ two
results from the coefficients of $u_0W_1^\mu\d_\mu\Phi_2$ and $u_2A_\mu
W_1^\mu$:
$$b_4=-m_2\sin\Theta=-{m_1^2\over m_2}\sin\Theta.$$
This implies
$$m_1=m_2=m_W,\eqno(3.15)$$
which is called W-mass from now on. The resulting coupling in the sector
(0,1,2) becomes
$$T_1^{012}=ig\Bigl\{\sin\Theta A^\nu(\Phi_2\d_\nu\Phi_1-\Phi_1\d_\nu
\Phi_2)$$ 
$$+m_W\sin\Theta A^\nu(W_{2\nu}\Phi_1-W_{1\nu}\Phi_2)+m_W\sin\Theta
(\tilde u_1u_0\Phi_2-\tilde u_2u_0\Phi_1)$$
$$+a_5\Bigl[W_1^\nu(\Phi_2\d_\nu\Phi_0-\Phi_0\d_\nu\Phi_2)-
W_2^\nu(\Phi_0\d_\nu\Phi_1-\Phi_1\d_\nu\Phi_0)$$
$$+2m_WW_{1\nu}W_2^\nu\Phi_0-m_W(\tilde u_1u_2+\tilde u_2u_1)\Phi_0
-{m_H^2\over m_W}\Phi_0\Phi_1\Phi_2\Bigl]\Bigl\},\eqno(3.16)$$
where $m_H$ is the undetermined mass of $\Phi_0$, the 'Higgs' mass.

Regarding the sector (1,1,2) we note that
terms with two fields $a=1$ do not occur in (3.7). Therefore, the
condition of gauge invariance in this sector leads to homogeneous
equations and it turns out that they only have the trivial solution 
with all parameters equal
to 0. The same is obviously true in the sectors (1,1,3), (1,3,3), 
as well as in (1,2,2), (2,2,3) and (2,3,3).
\vskip 0.5cm
\subsection{Sectors (0,1,1), (0,2,2), (0,3,3)}
\vskip 0.5cm
Although these sectors have also two fields with the same $a$, so that
there are no corresponding terms in (3.7), the situation is different
because $a=0$ behaves differently. The general ansatz 
for the coupling reads as follows
$$T_1^{011}=ig\Bigl[\alpha_1A_\mu W_1^\mu\Phi_1+\alpha_2W_{1\mu}
W_1^\mu\Phi_0+\beta_1\Phi_0\Phi_1^2+$$
$$+\gamma_1\tilde u_0u_1\Phi_1+\gamma_2\tilde u_1u_0\Phi_1+\gamma_3
\tilde u_1u_1\Phi_0+$$
$$+\delta_1W_1^\mu(\Phi_0\d_\mu\Phi_1-\Phi_1\d_\mu\Phi_0)\Bigl].$$
The gauge variation after taking out the derivatives of the ghost fields
is equal to
$$d_QT_1^{011}=-g\d^\mu\Bigl[\alpha_1u_0W_{1\mu}\Phi_1+\alpha_1u_1
A_\mu\Phi_1$$
$$+2\alpha_2u_1W_{1\mu}\Phi_0+\delta_1u_1(\Phi_0\d_\mu\Phi_1-\Phi_1\d_\mu
\Phi_0)+\delta_1m_1u_1W_{1\mu}\Phi_0\Bigl]\eqno(3.17.1)$$
$$-g\Bigl[u_0\d_\mu W_1^\mu\Phi_1(-\alpha_1-\gamma_2)+u_1\d_\mu W_1^\mu
\Phi_0(-2\alpha_2-\gamma_3-\delta_1m_1)-\alpha_1u_0W_1^\mu\d_\mu\Phi_1$$
$$+u_1W_1^\mu\d_\mu\Phi_02(-\alpha_2-\delta_1m_1)+u_1\Phi_0\Phi_1
(2\beta_1m_1+\gamma_3m_1+\delta_1(m_1^2-m_H^2))$$
$$-\gamma_1\eta_1\d_\mu A^\mu\Phi_1-\gamma_2m_1\Phi_0\Phi_1^2\Bigl].
\eqno(3.17.2)$$
For gauge invariance, the coefficients in (3.17.2) must vanish which
implies
$$\alpha_1=\gamma_1=\gamma_2=0,$$
$$\alpha_2=-m_W\delta_1=-\gamma_3,\quad\beta_1={m_H^2\over 2m_W}\delta_1.
$$

The results for the other two sectors are simply obtained by replacing
$a=1$ by 2 or 3:
$$T_1^{022}=ig\delta_2\Bigl[W_2^\mu(\Phi_0\d_\mu\Phi_2-\Phi_2\d_\mu
\Phi_0)-m_WW_{2\mu}W_2^\mu\Phi_0$$
$$+{m_H^2\over 2m_W}\Phi_0\Phi_2^2+m_W\tilde u_2u_2\Phi_0\Bigl]
\eqno(3.18)$$
$$T_1^{033}=ig\delta_3\Bigl[Z^\mu(\Phi_0\d_\mu\Phi_3-\Phi_3\d_\mu
\Phi_0)-m_zZ_\mu Z^\mu\Phi_0$$
$$+{m_H^2\over 2m_Z}\Phi_0\Phi_3^2+m_Z\tilde u_3u_3\Phi_0\Bigl].\eqno
(3.19)$$
\vskip 0.5cm
\subsection{Remaining Sectors}
\vskip 0.5cm

The sectors (0,1,3) and (0,2,3) are similar to (0,1,2) (3.16).
The only difference is that the 'Higgs'-free part is trivial
because $f_{013}=0=f_{023}$.
The sectors (0,0,1), (0,0,2), (0,0,3), 
(1,1,1), (2,2,2) and (3,3,3) are also trivial. 
The last sector (0,0,0) is trivial up to a
possible coupling $\sim b\Phi_0^3$.

Summing up, the scalar couplings are fixed by first order gauge
invariance as follows:
$$T_1^\Phi=ig\Bigl\{\sin\Theta A^\nu(\Phi_2\d_\nu\Phi_1-\Phi_1\d_\nu\Phi_2)
\eqno(3.20.1)$$
$$+\Bigl(1-{m_Z^2\over 2m_W^2}\Bigl)\cos\Theta Z^\nu
(\Phi_2\d_\nu\Phi_1-\Phi_1\d_\nu\Phi_2)+{m_Z\over 2m_W}\cos\Theta W_1^\nu 
(\Phi_3\d_\nu\Phi_2-\Phi_2\d_\nu\Phi_3)\eqno(3.20.2)$$
$$-{m_Z\over 2m_W}\cos\Theta W_2^\nu(\Phi_3\d_\nu\Phi_1-\Phi_1\d_\nu\Phi_3) 
+m_W\sin\Theta A_\nu(W_2^\nu\Phi_1-W_1^\nu\Phi_2)\eqno(3.20.3)$$
$$+\Bigl(m_W-{m_Z^2\over m_W}\Bigl)\cos\Theta(W_{2\nu}Z^\nu\Phi_1-W_{1\nu}
Z^\nu\Phi_2)
+m_W\sin\Theta(\tilde u_1u_0\Phi_2-\tilde u_2u_0\Phi_1)\eqno(3.20.4)$$
$$+{m_Z\over 2}\cos\Theta(\tilde u_2u_1-\tilde u_1u_2)\Phi_3+{m_Z^2\over
2m_W}\cos\Theta\tilde u_3(u_2\Phi_1-u_1\Phi_2)\eqno(3.20.5)$$
$$+\Bigl({m_Z^2\over 2m_W}-m_W\Bigl)\cos\Theta(\tilde u_2u_3\Phi_1-
\tilde u_1u_3\Phi_2)\eqno(3.20.6)$$
$$+a_5\Bigl[W_1^\nu(\Phi_2\d_\nu\Phi_0-\Phi_0\d_\nu\Phi_2)-
W_2^\nu(\Phi_0\d_\nu\Phi_1-\Phi_1\d_\nu\Phi_0)\eqno(3.20.7)$$
$$+2m_WW_{1\nu}W_2^\nu\Phi_0-m_W(\tilde u_1u_2+\tilde u_2u_1)\Phi_0
-{m_H^2\over m_W}\Phi_0\Phi_1\Phi_2\Bigl]\eqno(3.20.8)$$
$$+a_6[2\to 3]+a_7[1\to 3]$$
$$+\delta_1\Bigl[W_1^\nu(\Phi_0\d_\nu\Phi_1-\Phi_1\d_\nu\Phi_0)
-m_WW_{1\nu}W_1^\nu\Phi_0\eqno(3.20.9)$$
$$+{m_H^2\over 2m_W}\Phi_0\Phi_1^2+m_W\tilde u_1u_1\Phi_0\Bigl]
+\delta_2\Bigl[1\rightarrow 2\Bigl]\eqno(3.20.10)$$
$$+\delta_3\Bigl[Z^\nu(\Phi_0\d_\nu\Phi_3-\Phi_3\d_\nu\Phi_0)
-m_ZZ_\nu Z^\nu\Phi_0\eqno(3.20.11)$$
$$+{m_H^2\over 2m_Z}\Phi_0\Phi_3^2+m_Z\tilde u_3u_3\Phi_0\Bigl]
+b\Phi_0^3\Bigl\}.\eqno(3.20.12)$$

We notice that all coupling constants have been fixed by first order
gauge invariance, apart from the couplings $a_5, a_6, a_7, \delta_1, \delta_2,
\delta_3$ of the physical scalar $\Phi_0$ (the 'Higgs' field). 
If we set these constants =0
the Higgs field is decoupled and superfluous, therefore, first order
gauge invariance does not require $\Phi_0$. But at second order we will
find $\delta_j\ne 0$ for $j=1,2,3$, so that one physical scalar field is
indispensable. 

\vskip 1cm
\section{Gauge Invariance at Second Order}
\vskip 0.5cm

The time ordered products $T_2(x,y),\,T_{2/1}(x,y)$ are uniquely determined 
for $x\not= y$ by causality:
$$T_2(x,y)=T_1(x)T_1(y),\quad T_{2/1}^\nu (x,y)=T_{1/1}^\nu (x)T_1(y)
\quad{\rm for}\> x\not\in\{y\}+\bar V^-\eqno(4.1a)$$
and 
$$T_2(x,y)=T_1(y)T_1(x),\quad T_{2/1}^\nu (x,y)=T_1(y)T_{1/1}^\nu (x)
\quad{\rm for}\> y\not\in\{x\}+\bar V^-.\eqno(4.1b)$$
If $(x-y)^2<0$ both formulas hold true and this is consistent because
Wick polynomials commute for space-like separation.
We apply Wick's theorem to these expressions. The normally ordered products 
of the free field operators are defined also for $x=y$. Hence, the extension
to the diagonal $x=y$ is problematic for the C-number distributions only.
However, in the tree diagrams of $T_2\vert_{x\not= y},\,T_{2/1}
\vert_{x\not= y}$ the C-number distributions are the Feynman propagators
with derivatives ($D_F(x-y),\,\d_\mu D_F(x-y),\,\d_\mu\d_\nu D_F(x-y)$)
and they extend trivially to $x=y$. In other words the Feynman propagators
provide a distinguished extension $T_2\vert^0_{\rm tree},\, 
T_{2/1}\vert^0_{\rm tree}$. The most general extension which is compatible
with renormalizability by power counting reads
$$T_2\vert_{\rm tree}(x,y)=T_2\vert^0_{\rm tree}(x,y)+N_{(2)}(x,y),\quad
T_{2/1}^\nu\vert_{\rm tree}(x,y)=T_{2/1}^\nu\vert^0_{\rm tree}(x,y)
+N_{(2/1)}^\nu(x,y),\eqno(4.2a)$$
where $N_{(2)},\,N_{(2/1)}$ have the form
$$N_{(2)}(x,y),\,N_{(2/1)}(x,y)=\sum_{j_1,\ldots,j_4}C_{j_1j_2j_3j_4}
\delta (x-y):B_{j_1}(x)B_{j_2}(x)B_{j_3}(y)B_{j_4}(y):,\eqno(4.2b)$$
$C_{j_1j_2j_3j_4}\in {\bf C}$ are constants and $B_j\in\{A,W_1,W_2,
Z,u_a,\tilde u_a,\Phi_a,\Phi_0\}$. 
$N_{(2)}$ represents additional couplings of four fields at second order. It
corresponds to the quartic terms in the interaction Lagrangian.

Gauge invariance (1.5) is satisfied for $x\not= y$ by induction: for instance
in the region $x\not\in\{y\}+\bar V^-$ we have 
$$d_QT_2(x,y)=d_QT_1(x)T_1(y)+T_1(x)d_QT_1(y)=$$
$$=i\d_\nu^x(T_{1/1}^\nu (x)T_1(y))+i\d_\nu^y(T_1(x)T_{1/1}^\nu (y))=$$
$$=i\d_\nu^x T_{2/1}^\nu (x,y)+i\d_\nu^y T_{2/2}^\nu (x,y),\eqno(4.3)$$
where we have used\footnote{This equality holds true for all $x\not= y$ and 
must be maintained in the extension.} $T_{2/2}^\nu (x,y)=T_{2/1}^\nu (y,x)$.
Hence gauge invariance can only be violated by local terms $\sim D\delta
(x-y)$ (where $D$ is a differential operator) and this holds true also in 
the inductive construction of the higher orders. But, due to the 
non-uniqueness of the extension to the total diagonal, we still have such 
local terms at disposal, namely $N_{(2)},\,N_{(2/1)}$ in the case of (4.3).
To prove gauge invariance at orders $n\geq 2$
means to show that one can choose the normalizations
such that gauge invariance holds also on the total diagonal.
In addition the normalizations must also satisfy the other symmetry 
requirements, e.g. Poincar\'e covariance, pseudounitarity $S(h)^*S(h)=
{\bf 1}=S(h)S(h)^*$ (for $h$ real-valued).

For the loop diagrams at second order we have checked that one can fulfil
gauge invariance by choosing suitable normalizations. Much more interesting
are the {\it tree} diagrams at second order, because in this case gauge
invariance determines the parameters in the Higgs-coupling which are still
free, the mass ratio ${m_Z\over m_W}$ and $N_{(2)}$ which corresponds to the 
quartic terms in the interaction Lagrangian. We classify the tree
terms in (4.3) in the following way:

- the terms $\sim D_F(x-y),\,\sim\d_\mu D_F(x-y),\,\sim\d_\mu\d_\nu D_F(x-y)$
($\mu$ and $\nu$ not contracted) and $\sim\d_\mu\d_\nu\d_\lambda D_F(x-y)$
(no pair of Lorentz indices contracted) are the {\it non-local terms};

-the terms $\sim\delta(x-y),\sim\d\delta(x-y)$ are called {\it local}.

Other terms (especially higher derivatives) do not appear. The non-local
terms cancel for $x\not= y$ by (4.3). Therefore, they cancel also for $x=y$.
Hence, we need to consider the local terms only. There are the following kinds
of local terms in (4.3):

- the 'normalization terms' $d_QN_{(2)}(x,y),\,\d_\nu^xN_{(2/1)}^\nu (x,y)$
and $\d_\nu^yN_{(2/2)}^\nu (x,y)$;

- the 'anomalies'\footnote{The ordinary axial anomalies are of the same kind, 
they appear in the third order triangle diagrams with axial vector 
couplings to fermions (see part II, Sect.4). The difference is that the
axial anomalies cannot be removed by finite renormalizations.}
which are generated by taking the divergences $\d_\nu^x T_{2/1}^\nu
\vert^0_{\rm tree}(x,y)$ and \break
$\d_\nu^y T_{2/2}^\nu\vert^0_{\rm tree} (x,y)$
due to the $\delta$-distribution in 
$$\d_\nu\d^\nu D_F(x-y)=-m^2D_F(x-y)+\delta(x-y).\eqno(4.4)$$

Note that $d_QT_2\vert^0_{\rm tree}(x,y)$ contains no local terms because 
$d_Q$ operates only on the field operators. To prove gauge invariance we 
must show that all anomalies can be removed by a suitable choice of
$N_{(2)}$ and $N_{(2/1)}$. (The latter determines $N_{(2/2)}$ by exchanging
$x\leftrightarrow y$.) The anomalies depend on the parameters which are 
still free in $T_1^\Phi$ (3.20) and on the mass ratio ${m_Z\over m_W}$.
It will turn out that the removal of the anomalies is only possible 
if these parameters and ${m_Z\over m_W}$ take certain values.

\vskip 0.5cm
\subsection{Simplification of the Scalar Coupling}
\vskip 0.5cm

There is only one way to generate an anomaly in $\d_\mu^x 
T_{2/1}^\mu\vert^0_{\rm tree}(x,y)\equiv\d_\mu^x T[T_{1/1}^\mu (x)T_1(y)]
\vert^0_{\rm tree}$: the corresponding term in $T_{1/1}^\mu (x)$ contains
a field operator $\d^\mu\varphi(x)\>(\varphi =A^\mu,\,W_1^\mu,\,W_2^\mu,\,
Z^mu,\,u_a,\,\tilde u_a,\break
\Phi_a,\,\Phi_0)$ and the latter is contracted with a field
operator in $T_1(y)$, i.e. there is a Feynman propagator $\d^\mu D_F(x-y)$.
Taking then the divergence $\d_\mu^x$ a term $\sim\delta (x-y)$ is produced
due to (4.4). Hence we are interested in those terms in $T^\mu_{1/1}$
which contain a derivative $\d^\mu$ with the same Lorentz index $\mu$. 
Examining the divergence terms in
the results of Sect.3, we collect the following list of terms
that generate anomalies:
$$T^\mu_{1/1}\vert_{\rm an}=ig\Bigl\{\sin\Theta (u_1W_2^\nu-u_2W_1^\nu) 
\d^\mu A_\nu \eqno(4.5.1)$$
$$+\sin\Theta (u_2A^\nu-u_0W_2^\nu)\d^\mu W_{1\nu}+\sin\Theta (u_0
W_1^\nu-u_1A^\nu)\d^\mu W_{2\nu} \eqno(4.5.2)$$
$$+\cos\Theta\Bigl[(u_2Z^\nu-u_3W_2^\nu)\d^\mu W_{1\nu}
+(u_3W_1^\nu-u_1Z^\nu)\d^\mu W_{2\nu}\eqno(4.5.3)$$
$$+(u_1W_2^\nu-u_2W_1^\nu)\d^\mu Z_\nu\Bigl]\eqno(4.5.4)$$ 
$$+\sin\Theta (u_0u_1\d^\mu\tilde u_2+u_2u_0\d^\mu\tilde u_1+ 
u_1u_2\d^\mu\tilde u_0)\eqno(4.5.5)$$
$$+\cos\Theta (u_2u_3\d^\mu\tilde u_1+u_3u_1\d^\mu\tilde u_2 
+u_1u_2\d^\mu\tilde u_3)\eqno(4.5.6)$$ 
$$+\sin\Theta u_0(\Phi_2\d^\mu\Phi_1-\Phi_1\d^\mu\Phi_2)+\Bigl(1-{m_Z^2
\over 2m_W^2}\Bigl)\cos\Theta
u_3(\Phi_2\d^\mu\Phi_1-\Phi_1\d^\mu\Phi_2)\eqno(4.5.7)$$ 
$$+{m_Z\over 2m_W}\cos\Theta\Bigl[(u_2\Phi_1-u_1\Phi_2)\d^\mu\Phi_3+u_1
\Phi_3\d^\mu\Phi_2-u_2\Phi_3\d^\mu\Phi_1\Bigl]
\eqno(4.5.8)$$
$$+a_5[u_1(\Phi_2\d^\mu\Phi_0-\Phi_0\d^\mu\Phi_2)
+u_2(\Phi_1\d^\mu\Phi_0-\Phi_0\d^\mu\Phi_1)]+a_6[\ldots]+a_7[\ldots] 
\eqno(4.5.9)$$
$$+\delta_1u_1(\Phi_0\d^\mu\Phi_1-\Phi_1\d^\mu\Phi_0)
+\delta_2u_2(\Phi_0\d^\mu\Phi_2-\Phi_2\d^\mu\Phi_0)\eqno(4.5.10)$$
$$+\delta_3u_3(\Phi_0\d^\mu\Phi_3-\Phi_3\d^\mu\Phi_0)\Bigl\}.\eqno(4.5.11)$$

Let us now consider the combination of the second term in (4.5.9) with
the first term in (3.20.1). There comes an anomaly from the contraction
of $\d^\mu\Phi_2(x)$ with $\Phi_2(y)$. We denote the corresponding term by
$$g^2\sin\Theta a_5u_1(x)\Phi_0(x)D_F[\d^\mu\Phi_2(x),\Phi_2(y)]
A^\nu(y)\d_\nu\Phi_1(y),\eqno(4.6)$$
where $D_F[\d^\mu\Phi_2(x),\Phi_2(y)]=-i\d^\mu D_F(x-y)$.
It results the anomaly
$$A_1=-ig^2\sin\Theta a_5u_1\Phi_0A^\nu\d_\nu\Phi_1\delta(x-y).\eqno(4.7)$$

The same second term in (4.5.9) with the second one in
(3.20.1) gives rise to another anomaly
$$\d_\mu^x T_{2/1}^\mu\vert^0_{\rm tree}(x,y)=\d_\mu^x\{
-g^2\sin\Theta a_5u_1(x)\Phi_0(x)D_F[\d^\mu\Phi_2(x),\d^\nu\Phi_2(y)]
A_\nu(y)\Phi_1(y)\}+...=$$
$$=ig^2\sin\Theta a_5 u_1(x)\Phi_0(x)\Phi_1(y)
A^\nu(y)\d_\nu^y\delta(x-y)+...\>.
\eqno(4.8)$$
We liberate the $\delta$-distribution from the derivative by means of
$$A(x)B(y)\d_\mu^x\delta(x-y)+A(y)B(x)\d_\mu^y\delta(x-y)=$$
$$=A(x)(\d_\mu B)(x)\delta(x-y)-(\d_\mu A)(x)B(x)\delta(x-y).\eqno(4.9)$$
Here we have added the other anomaly with $x$ and $y$ interchanged which
comes from \break $\d_\mu^y T_{2/2}^\mu\vert^0_{\rm tree}(x,y)$. There is a 
contribution from $\d_\mu^x N^\mu_{2/1}(x,y)$ which belongs to (4.8)
\footnote{The most general Poincar\'e covariant extension (to the 
diagonal $x=y$) of 
$D_F[\d^\mu\Phi_2(x),\d^\nu\Phi_2(y)]=i\d^\mu\d^\nu D_F(x-y)$ (4.8)
which is compatible with renormalizability by power 
counting reads
$$i\d^\mu\d^\nu D_F(x-y)+\alpha_1 g^{\mu\nu}\delta(x-y)$$
where $\alpha_1$ is an arbitrary constant. In (4.10) we consider the
$\alpha_1$-part.}
$$\d_\mu^x N^\mu_{2/1}(x,y)=ig^2\sin\Theta a_5\alpha_1\d_\nu^y
\Bigl(u_1(x)\Phi_0(x)\Phi_1(y)A^\nu(y)\delta (x-y)\Bigl),\eqno(4.10)$$
where $\alpha_1$ is a free parameter. Again we add the term with $x\lra y$
(coming from $\d_\mu^y N^\mu_{2/2}(x,y)$) and use the relation
$$\d_\mu^x[A(x)B(y)\delta(x-y)]+\d_\mu^y[A(y)B(x)\delta(x-y)]=$$
$$=(\d_\mu A)(x)B(x)\delta(x-y)+A(x)(\d_\mu B)(x)\delta(x-y).\eqno(4.11)$$
Summing up we obtain from (4.8), (4.10) and the corresponding expressions
with $x\lra y$ the local terms
$$A_2=ig^2\sin\Theta a_5:\Bigl[(1+\alpha_1)\d_\nu u_1
\Phi_0A^\nu\Phi_1+(1+\alpha_1)u_1\d_\nu\Phi_0A^\nu\Phi_1$$
$$+(\alpha_1-1)u_1\Phi_0\d_\nu A^\nu\Phi_1+(\alpha_1-1)u_1\Phi_0
A^\nu\d_\nu\Phi_1\Bigl]:\delta(x-y).\eqno(4.12)$$

Let us now collect all terms with field operators $u_1\Phi_0A^\nu\d_\nu
\Phi_1$. Since the anomaly $A_1$ (4.7) gets a factor 2 if the term with
$x$ and $y$ interchanged is included, the result is proportional to
$(3-\alpha_1)a_5$. There is another anomaly $A_3$ with the same external
field operators coming from the forth term in (4.5.2) contracted with
the third one in (3.20.7) which has the same value as $A_1$, i.e. $A_3=A_1$. 
There is no contribution from the normalization term $d_QN_{(2)}$ of 
$d_QT_2\vert_{\rm tree}$ (4.2). This is due to the external field 
operator $\d_\nu\Phi_1$, 
because the derivatives on external field operators in $d_QN_{(2)}$ (4.2b)
must come from $d_Q$. Hence, a gauge invariant theory requires 
$$(5-\alpha_1)a_5=0.$$
Repeating the same argument for the terms $\sim
u_1\Phi_1A^\nu\d_\nu\Phi_0$ leads to
$$(3+\alpha_1)a_5=0.$$
Both equations together yield $a_5=0$.
In the same way one finds $a_6=0=a_7$.

\vskip 0.5cm
\subsection{Determination of the Remaining Parameters}
\vskip 0.5cm
As at first order (Sect.3.2) we discuss gauge invariance in different
sectors specified by the indices $a$ of the external field operators.
Since the physical scalar $\Phi_0$ plays a special role (it is not 
affected by
$d_Q$, nor is it the result of a $d_QX$), the sectors can be further
subdivided, depending on the number of $\Phi_0$'s that occur. Let us first
turn to the sector $(\Phi_0,0,1,2)$. Here, contracting the second term in
(4.5.10) with the second one in (3.20.1) we get the anomaly
$$A_1=-ig^2\sin^2\Theta\delta_1u_1\Phi_0A^\nu(\d_\nu
\Phi_2)\delta(x-y).\eqno(4.13)$$
As in (4.12), the same term in (4.5.10) combined with the first term 
in  (3.20.1) gives
$$A_2=-g^2\sin\Theta\delta_1\Bigl[(1+\alpha_2)\d_\nu u_1\Phi_0 
A^\nu\Phi_2+(1+\alpha_2)u_1\d_\nu\Phi_0A^\nu \Phi_2$$
$$+(\alpha_2-1)u_1\Phi_0\d_\nu A^\nu\Phi_2 
+(\alpha_2-1)u_1\Phi_0 A^\nu\d_\nu\Phi_2\Bigl]\delta(x-y),\eqno(4.14)$$
where $\alpha_2$ is a new free normalization constant.
Further anomalies come from the last term in (4.5.2) contracted with
the last one in (3.20.10). The latter actually consists of five terms
obtained from the square bracket in (3.20.9-10) by substituting 1 by 2
everywhere. The first and second of those five terms yield
$$A_3=-g^2\sin\Theta\delta_2 u_1A^\nu(\Phi_0\d_\nu\Phi_2-\d_\nu\Phi_0
\Phi_2)\delta(x-y).\eqno(4.15)$$
Since $A_1$ and $A_3$ get multiplied by 2, we obtain the following two
relations from setting the coefficients of $u_1\Phi_0A^\nu\d_\nu\Phi_2$
and $u_1\d_\nu\Phi_0A^\nu\Phi_2$ equal to zero:
$$3\delta_1-\alpha_2\delta_1-2\delta_2=0,\quad -(1+\alpha_2)\delta_1
+2\delta_2=0.$$
This implies
$$\delta_2=\delta_1.\eqno(4.16)$$

We proceed in the same way in the sector $(\Phi_0,1,2,3)$. From the
coefficients of $u_3W_1^\nu\Phi_0\d_\nu\Phi_2$ and $u_3W_1^\nu\Phi_2\d
_\nu\Phi_0$ we find
$$\delta_3=\delta_1{m_Z\over m_W}.\eqno(4.17)$$
Next, in the sector (2,2,3,3) we get from the coefficients of
$u_2Z^\nu\d_\nu\Phi_2\Phi_3$ and $u_2Z^\nu\Phi_2\d_\nu\Phi_3$ the result
$$4\delta_2\delta_3={m_Z^3\over m_W^3}\cos^2\Theta.$$ 
Inserting (4.16) and (4.17) we arrive at
$$\delta_1=\pm{m_Z\over 2m_W}\cos\Theta,\eqno(4.18)$$
so that all coupling constants are now determined, apart from $b$ in
(3.20.12). The solution is
unique up to the sign (4.18) of the 'Higgs' couplings. We note that the
latter are different from zero, so that $\Phi_0$ is really necessary for
gauge invariance.

Finally, in the sector (1,1,2,2) we obtain another result from the
coefficients of $u_1W_2^\nu\Phi_2$ $\d_\nu\Phi_1$ and
$u_1W_2^\nu\Phi_1\d_\nu\Phi_2$:
$$\delta_1\delta_2=1-3{m_Z^2\over 4m_W^2}\cos^2\Theta=\delta_1^2.
\eqno(4.19)$$
This is compatible with (4.18) if and only if
$$m_W=m_Z\cos\Theta,\eqno(4.20)$$
assuming $\cos\Theta>0$. This is a
direct consequence of gauge invariance. The final
values of the coupling constants of the physical scalar are now given by
$$\delta_1=\delta_2={\eps_1\over 2},\quad \delta_3={\eps_1\over 2
\cos\Theta},\quad \eps_1=\pm 1.\eqno(4.21)$$
The sign $\eps_1$ can be absorbed by a redefinition of $\Phi_0$. 
Then all couplings are in
precise agreement with what is obtained from the standard model after
spontaneous symmetry breaking \cite{13}.
To verify gauge invariance completely for second order tree diagrams, 
we have to show that with
the above values of the parameters all anomalies cancel out with proper
choice of the normalization constants. This is done in the appendix. 
\vskip 1cm
\section{Coupling to Leptons}
\vskip 1cm
To determine the coupling to leptons by the same method as before, we start
from the following ansatz
$$T_1^F=ig\Bigl\{b_1W_\mu^+\oe\gamma^\mu\nu+b'_1W_\mu^+\oe\gamma^\mu 
\gamma^5\nu$$
$$+b_2W_\mu^-\onu\gamma^\mu{\rm e}+b'_2W_\mu^-\onu\gamma^\mu\gamma^5{\rm e}
$$
$$+b_3Z_\mu\oe\gamma^\mu{\rm e}+b'_3Z_\mu\oe\gamma^\mu\gamma^5{\rm e}
$$
$$+b_4Z_\mu\onu\gamma^\mu\nu+b'_4Z_\mu\onu\gamma^\mu\gamma^5\nu
$$
$$+b_5A_\mu\oe\gamma^\mu{\rm e}+b'_5A_\mu\oe\gamma^\mu\gamma^5{\rm e}$$
$$+c_1\Phi^+\oe\nu+c'_1\Phi^+\oe\gamma^5\nu
+c_2\Phi^-\onu{\rm e}+c'_2\Phi^-\onu\gamma^5{\rm e}$$
$$+c_3\Phi_3\oe{\rm e}+c'_3\Phi_3\oe\gamma^5{\rm e}
+c_4\Phi_3\onu\nu+c'_4\Phi_3\onu\gamma^5\nu$$
$$+c_0\Phi_0\oe{\rm e}+c'_0\Phi_0\oe\gamma^5{\rm e}
+c_5\Phi_0\onu\nu+c'_5\Phi_0\onu\gamma^5\nu\Bigl\},\eqno(5.1)$$
where we have used the usual definitions $\Phi^\pm=(\Phi_1\pm i\Phi_2) 
/\sqrt{2}$,
and similarly for $W_\mu^\pm$ and $u^\pm$. Here we have assumed the
usual electric charges of the particles and charge conservation in each
term. In particular there is no coupling of the photon to the neutrini. 
A more general situation is considered in part II.
For simplicity we only
consider one family of leptons, the 'electron' and the 'neutrino',
which have arbitrary masses and fulfill the Dirac equations
$$\ds{\rm e}=-im_e{\rm e},\quad \d_\mu(\oe\gamma^\mu)=im_e\oe$$
$$\ds\nu=-im_\nu\nu,\quad \d_\mu(\onu\gamma^\mu)=im_\nu\nu.\eqno(5.2)$$
{\it We do not assume chiral fermions in {\rm (5.1)}, instead, we will get them
out as a consequence of gauge invariance at second order.} Of course,
this small theory is not gauge invariant at third order due to the
axial anomalies. As usual, this defect can be removed by adding the
quark degrees of freedom (see part II).

Since the fermions are not transformed by $d_Q$, first order gauge
invariance 
$$d_QT_1^F={\rm divergence}$$
immediately gives $b'_5=0$ assuming the electron mass
to be non-vanishing. This fact that the photon has no
axial-vector coupling can be traced back to the absence of a scalar
partner for the photon, that means to its vanishing mass. The massive 
gauge fields must
have axial-vector couplings. The other coupling constants are restricted at
first order as follows
$$c_1=i{m_e-m_\nu\over m_W}b_1,\quad c'_1=i{m_e+m_\nu\over m_W}b'_1$$
$$c_2=i{m_\nu-m_e\over m_W}b_2,\quad c'_2=i{m_e+m_\nu\over m_W}b'_2$$
$$c_3=0,\quad\quad c'_3=2i{m_e\over m_Z}b'_3$$
$$c_4=0,\quad\quad c'_4=2i{m_\nu\over m_Z}b'_4.\eqno(5.3)$$
The
resulting divergence form of $d_QT_1^F$ gives the following Q-vertex:
$d_QT_1^F=i\d_\mu T_{1/1}^{F\mu}$ with
$$T_{1/1}^{F\mu}=ig\{b_1u^+\oe\gamma^\mu\nu+b'_1u^+\oe\gamma^\mu\gamma^5
\nu$$
$$+b_2u^-\onu\gamma^\mu{\rm e}+b'_2u^-\onu\gamma^\mu\gamma^5
{\rm e}+b_3u_3\oe\gamma^\mu{\rm e}+b'_3u_3\oe\gamma^\mu\gamma^5
{\rm e}$$
$$+b_4u_3\onu\gamma^\mu\nu+b'_4u_3\onu\gamma^\mu\gamma^5\nu+b_5u_0
\oe\gamma^\mu{\rm e}\}.\eqno(5.4)$$

In the discussion of gauge invariance of second order tree graphs there
is a slight modification. While the anomalies coming from contractions
between $T_{1/1}^\mu$ (4.5) and $T_1^F$ can be calculated as before,
there is a new source of anomalies when we contract $T_{1/1}^{F\mu}$
with $T_1^F$. The resulting fermionic contractions $\sim S_m^F(x-y)$
give rise to anomalies if only one derivative is
applied:
$$i\d_\mu^x\gamma^\mu S_m^F(x-y)=mS_m^F(x-y)+\delta(x-y).\eqno(5.5)$$
In fact there is a derivative because we take the divergence with respect 
to the $Q$-vertex in e.g. $\d_\mu^x T_{2/1}^\mu\vert^0_{\rm tree}(x,y)$. Hence
every Fermi field in (5.4) generates an anomaly if it is contracted with
another Fermi field in $T_1^F$. But note that the
contractions of $T^F_{1/1}$ (5.4) with $T_1$ (3.20) do not produce any
anomaly. Things are greatly simplified by the fact that the
normalization terms $d_QN_{(2)}(x,y),\,\d_\nu^xN_{(2/1)}^\nu (x,y)$
and $\d_\nu^yN_{(2/2)}^\nu (x,y)$ have no contributions
with fermionic field operators. This is due to the fact that the spinor
fields $\psi$ and $\bar\psi$ have mass dimension ${3\over 2}$ (instead of the 
value $1$ of the other fields). Hence a term $\sim\delta(x-y)
:B_1B_2\bar\psi\psi:$ in $N_{(2)}$, $N_{(2/1)}$ (4.2b) 
would violate renormalizability
by power counting. In addition, dim$\,\psi={3\over 2}$ implies also that
there are no terms $\sim\d_\mu\delta(x-y)$ in the leptonic sectors of 
(4.3). In the sum of
all anomalies the coefficient of every Wick monomial must add up to 0 
in order to have gauge invariance.

This requirement determines
the parameters in (5.1) which are still free as follows. Assuming $b_1,
b_2\ne 0$, we find from the coefficients of $u_3\Phi_3\oe\gamma^5{\rm e}$
and $u_3\Phi_3\onu\gamma^5\nu$
$$c'_0=0,\quad c'_5=0,\eqno(5.6)$$
and from $u^+A_\nu\oe\gamma^\nu\nu$ the electric charge
$$gb_5=g\sin\Theta.\eqno(5.7)$$
Then, from $u_3\Phi_3\oe{\rm e}$ and $u_3\Phi_0\oe\gamma^5{\rm e}$
we get the two relations
$$2b'_3c'_3={ic_0\over 2\cos\Theta},\quad
2b'_3c_0=-{ic'_3\over 2\cos\Theta}.$$
By (5.3) this leads to
$$c_0={m_e\over 2m_W}\eqno(5.8)$$
$$b'_3={\eps_2\over 4\cos\Theta},\quad\eps_2=\pm 1.\eqno(5.9)$$
A possible trivial solution $c_0=0=b'_3$ is excluded by later conditions.
Similarly we find from $u_3\Phi_3\onu\nu$ and $u_3\Phi_0\onu\gamma^5\nu$
$$2b'_4c'_4={ic_5\over 2\cos\Theta},\quad
2b'_4c_5=-{ic'_4\over 2\cos\Theta},\eqno(5.10)$$
which, with (5.3), yields
$$c_5={m_\nu\over 2m_W}\eqno(5.11)$$
$$2b'_4={\eps_3\over 4\cos\Theta}.\eqno(5.12)$$
>From $u_1W_2^\mu\oe\gamma_\mu\gamma^5{\rm e}$ and
$u_1W_2^\mu\onu\gamma_\mu\gamma^5\nu$ we obtain
$$b_1b'_2+b'_1b_2=b'_3\cos\Theta=-b'_4\cos\Theta,\eqno(5.13)$$
which determines the sign $\eps_3$ in (5.12):
$$b'_3=-b'_4={\eps_2\over 4\cos\Theta}.\eqno(5.14)$$

>From $u^+\Phi_3\oe\nu$ we find
$$c_1=2b'_1(c'_3+c'_4).$$
If we use (5.3) on both sides we arrive at
$$b_1=\eps_2b'_1.\eqno(5.15)$$
The same reasoning with $u^-\Phi_3\onu{\rm e}$ gives
$$b_2=\eps_2b'_2.\eqno(5.16)$$
Substituting these results into (5.13) we get
$$b_1b_2={1\over 8}=b'_1b'_2.\eqno(5.17)$$
>From $u_1W_2^\mu\onu\gamma_\mu\nu$ we now find
$$b_4\cos\Theta=-b_1b_2-b'_1b'_2=-{1\over 4}.\eqno(5.18)$$
Finally, $u_1W_2^\mu\oe\gamma_\mu{\rm e}$ gives the relation
$$b_3\cos\Theta=b_1b_2+b'_1b'_2-\sin\Theta b_5,$$
which determines
$$b_3={1\over 4\cos\Theta}-\sin\Theta\tan\Theta.\eqno(5.19)$$

Now we are ready to write down the leptonic coupling, as far as it is
restricted by gauge invariance alone:
$$T_1^F=ig\biggl\{b_1W_\mu^+\oe\gamma^\mu(1+\eps_2\gamma^5)\nu
+b_2W_\mu^-\onu\gamma^\mu(1+\eps_2\gamma^5){\rm e}$$
$$+{1\over 4\cos\Theta}Z_\mu\oe\gamma^\mu(1+\eps_2\gamma^5){\rm e}
-\sin\Theta\tan\Theta Z_\mu\oe\gamma^\mu{\rm e}$$
$$-{1\over 4\cos\Theta}Z_\mu\onu\gamma^\mu(1+\eps_2\gamma^5)\nu
+\sin\Theta A_\mu\oe\gamma^\mu{\rm e}$$
$$+i{m_e-m_\nu\over m_W}b_1\Phi^+\oe\nu
+i{m_e+m_\nu\over m_W}b_1\eps_2\Phi^+\oe\gamma^5\nu$$
$$-i{m_e-m_\nu\over m_W}b_2\Phi^-\onu{\rm e}
+i{m_e+m_\nu\over m_W}b_2\eps_2\Phi^-\onu\gamma^5{\rm e}$$
$$+{i\eps_2\over 2m_W}m_e\Phi_3\oe\gamma^5{\rm e}
-i\eps_2{m_\nu\over 2m_W}\Phi_3\onu\gamma^5\nu$$
$$+{m_e\over 2m_W}\Phi_0\oe{\rm e}
+{m_\nu\over 2m_W}\Phi_0\onu\nu\biggl\}.\eqno(5.20)$$
We have verified that with these parameters all further conditions for 
other Wick monomials are satisfied. This completes the proof of gauge
invariance for second order tree graphs. The importance of the result (5.20) lies in the
chiral coupling $\sim (1+\eps_2\gamma^5)$ of the fermions. The sign
$\eps_2$ is conventional. We see that perturbative gauge invariance is
the origin of maximal parity
violation in weak interactions and for the universality of the
couplings (i.e. there is only one independent coupling constant). 
All couplings are in agreement with the standard model \cite{13}. 
$b_1$ and $b_2$ can be further restricted as follows: Pseudounitarity 
$S(h)^*S(h)={\bf 1}=S(h)S(h)^*$ (for $h$ real-valued)
requires $b_1^*=b_2$ \cite{7}; absorbing the phase of $b_1$ by a
redefinition of the field operator ${\rm e}(x)$ we have 
$b_1=b_2$. Inserting this into (5.17) we get the usual values
$$b_1=b_2={1\over 2\sqrt{2}}.\eqno(5.21)$$

\vskip 0.5cm
\section*{Appendix: Verification of gauge invariance}
\vskip 0.5cm
With the techniques described in Sect.4 it is not hard to verify that
the very many anomalies all cancel out, if the normalization terms are
chosen properly. The normalization terms $N_{(2)}$ of $T_2\vert_{\rm tree}$ 
(4.2) are of particular interest because they represent additional
couplings of four fields at second order which are required by gauge
invariance. We list them all below.

The anomalies without external scalar fields are exactly the same as in
the massless case because there are no anomalies of this type coming
from scalar contractions. The compensation of these anomalies has been 
shown for arbitrary
Yang-Mills theories in \cite{3}, hence, we need not consider them here again.
The corresponding normalization terms (which are part of $N_{(2)}$ (4.2))
are the four-boson interactions
and agree precisely with the standard model. 

There exists a symmetry in the indices 1 and 2: terms without
$\Phi_0$ in (3.20) and (4.5) are antisymmetric under exchange of
$1\leftrightarrow 2$, terms with $\Phi_0$ are symmetric. 
Consequently, every anomaly has a
symmetry-partner with the same scheme of compensation. We only list one
of the two partners below. An anomaly is specified by the term in (4.5)
which is contracted at $\d^\mu X$ with a certain term in (3.20). The
terms are identified as follows: (10/1-9/2) means the 1st term in
(4.5.10) contracted with the 2nd one in (3.20.9) etc.; the notion
(3.20.10/3) represents the 5 terms in the last square bracket in
(3.20.10) which are not explicitly written down. From the resulting
anomalies we collect the contributions with the listed field operators,
their cancellation takes place according to the following list:

\noindent
{\it 1) Sector $(\Phi_0,\Phi_0,1,1)$}:

$u_1W_1^\nu\Phi_0\d_\nu\Phi_0:\> (10/1-9/2)+(10/1-9/1)=0$

$u_1\d_\nu W_1^\nu\Phi_0^2:\> (10/1-9/1)=0$

$\d_\nu u_1W_1^\nu\Phi_0^2:\> (10/1-9/1)=d_QN_1$

$u_1\Phi_0^2\Phi_1:\> (10/2-12/3)+(10/1-10/1)=d_QN_2$
$$N_1={i\over 4}g^2(W_{1\nu}W_1^\nu+W_{2\nu}W_2^\nu)\Phi_0^2 
\delta(x-y)\eqno(A.1)$$
$$N_2=ig^2\Bigl({m_H^2\over 4m_W^2}-{3b\over 2m_W}\Bigl)\Phi_0^2 
(\Phi_1^2+\Phi_2^2)\delta(x-y)\eqno(A.2)$$
{\it 2) Sector $(\Phi_0,\Phi_0,3,3)$}:

$u_3Z^\nu\Phi_0\d_\nu\Phi_0:\> (11/1-11/2)+(11/1-11/1)=0$

$u_3\d_\nu Z^\nu\Phi_0^2:\> (11/1-11/1)=0$

$\d_\nu u_3 Z^\nu\Phi_0^2:\> (11/1-11/1)=d_QN_3$

$u_3\Phi_0^2\Phi_3:\> (11/2-12/3)+(11/1-12/1)=d_QN_4$
$$N_3={ig^2\over 4\cos^2\Theta}Z_\nu Z^\nu\Phi_0^2\delta(x-y)
\eqno(A.3)$$
$$N_4=ig^2\Bigl({m_H^2\over 4m_W^2}-{3b\over 2m_W}\Bigl) 
\Phi_0^2\Phi_3^2\delta(x-y)\eqno(A.4)$$
{\it 3) Sector $(\Phi_0,0,1,2)$}:

$u_1A^\nu\Phi_0\d_\nu\Phi_2:\> (10/1-1/2)+(10/1-1/1)+(2/4-10/3)=0$

$u_1A^\nu\d_\nu\Phi_0\Phi_2:\> (10/1-1/1)+(2/4-10/3)=0$

$u_1\d_\nu A^\nu\Phi_0\Phi_2:\> (10/1-1/1)=0$

$\d_\nu u_1A^\nu\Phi_0\Phi_2:\> (10/1-1/1)=d_QN_5$

$u_2A^\nu W_{1\nu}\Phi_0:\> (10/3-3/4)+(2/1-9/3)=d_QN_5$

$u_0W_1^\nu\Phi_0\d_\nu\Phi_2:\> (7/1-9/1)+(2/3-10/3)=0$

$u_0W_1^\nu\d_\nu\Phi_0\Phi_2:\> (7/1-9/1)+(2/3-10/3)+(7/1-9/2)=0$

$u_0\d_\nu W_1^\nu\Phi_0\Phi_2:\> (7/1-9/1)=0$

$\d_\nu u_0W_1^\nu\Phi_0\Phi_2:\> (7/1-9/1)=d_QN_5$

$\tilde u_2u_1u_0\Phi_0:\> (10/1-4/4)+(5/1-10/3)=0$

$u_0\Phi_0\Phi_1\Phi_2:\> (7/1-10/1)+(7/2-10/3)=0$
$$N_5=ig^2\sin\Theta \Phi_0A_\nu(W_1^\nu\Phi_2 
-W_2^\nu\Phi_1)\delta(x-y)\eqno(A.5)$$
{\it 4) Sector $(\Phi_0,1,2,3)$}:

$u_3W_1^\nu\Phi_0\d_\nu\Phi_2:\> (3/3-10/3)+(7/3-9/1)+(11/1-2/3)+(11/1-2/4)=0$

$u_3W_1^\nu\d_\nu\Phi_0\Phi_2:\> (3/3-10/3)+(7/3-9/1)+(7/3-9/2)+(11/1-2/4)=0$

$u_3\d_\nu W_1^\nu\Phi_0\Phi_2:\> (11/1-2/4)+(7/3-9/1)=0$

$\d_\nu u_3W_1^\nu\Phi_0\Phi_2:\> (11/1-2/4)+(7/3-9/1)=d_QN_6$

$u_2W_1^\nu Z_\nu\Phi_0:\> (3/1-9/3)+(4/2-11/3)+(10/3-4/2)=d_QN_6$

$u_1Z^\nu\Phi_0\d_\nu\Phi_2:\> (3/4-10/3)+(8/2-11/1)+(10/1-2/2)+(10/1-2/1)=0$

$u_1Z^\nu\d_\nu\Phi_0\Phi_2:\> (10/1-2/1)+(8/2-11/1)+(3/4-10/3)+(8/2-11/2)=0$

$u_1\d_\nu Z^\nu\Phi_0\Phi_2:\> (10/1-2/1)+(8/2-11/1)=0$

$\d_\nu u_1Z^\nu\Phi_0\Phi_2:\> (10/1-2/1)+(8/2-11/1)=d_QN_6$

$u_3\Phi_0\Phi_1\Phi_2:\> (7/3-10/1)+(7/4-10/3)=0$

$\tilde u_1u_2u_3\Phi_0:\> (10/3-6/2)+(6/1-10/2)+(11/1-5/2)=0$

$\tilde u_3u_1u_2\Phi_0:\> (10/1-5/3)+(6/3-12/2)=0$

$u_2\Phi_0\Phi_1\Phi_3:\> (8/1-12/1)+(8/4-10/1)=0$

$u_2W_1^\nu\Phi_0\d_\nu\Phi_3:\> (4/2-11/1)+(8/4-9/1)+(10/3-2/3)+(10/3-2/4)=0$

$u_2W_1^\nu\d_\nu\Phi_0\Phi_3:\> (4/2-11/2)+(8/4-9/1)+(8/4-9/2)+(10/3-2/3)=0$

$u_2\d_\nu W_1^\nu\Phi_0\Phi_3:\> (10/3-2/3)+(8/4-9/1)=0$

$\d_\nu u_2W_1^\nu\Phi_0\Phi_3:\> (10/3-2/3)+(8/4-9/1)=0$
$$N_6=-ig^2\sin\Theta\tan\Theta Z_\nu\Phi_0(W_1^\nu \Phi_2-W_2^\nu
\Phi_1)\delta(x-y)\eqno(A.6)$$
{\it 5) Sector $(0,0,1,1)$}:

$u_0A^\nu\Phi_1\d_\nu\Phi_1:\> (7/2-1/1)+(7/2-1/2)=0$

$u_0\d_\nu A^\nu\Phi_1^2:\> (7/2-1/2)=0$

$\d_\nu u_0A^\nu\Phi_1^2:\> (7/2-1/2)=d_QN_7$

$u_1A_\nu A^\nu\Phi_1:\> (2/4-3/3)=d_QN_7$

$u_0A_\nu W_1^\nu\Phi_1:\> (2/3-3/3)+(7/2-3/4)=0$
$$N_7=ig^2\sin^2\Theta A_\nu A^\nu(\Phi_1^2+\Phi_2^2)\delta(x-y)\eqno(A.7)$$
{\it 6) Sector (0,1,1,3)}:

$u_0W_1^\nu\Phi_1\d_\nu\Phi_3:\> (2/3-3/2)+(7/2-2/3)+(7/2-2/4)=0$

$u_0W_1^\nu\d_\nu\Phi_1\Phi_3:\> (2/3-3/1)+(7/2-2/3)=0$

$u_0\d_\nu W_1^\nu\Phi_1\Phi_3:\> (7/2-2/3)=0$

$\d_\nu u_0W_1^\nu\Phi_1\Phi_3:\> (7/2-2/3)=d_QN_8$

$u_1A^\nu\Phi_1\d_\nu\Phi_3:\> (2/4-3/2)+(8/3-1/2)=0$

$u_1A^\nu\d_\nu\Phi_1\Phi_3:\> (2/4-3/1)+(8/3-1/1)+(8/3-1/2)=0$

$u_1\d_\nu A^\nu\Phi_1\Phi_3:\> (8/3-1/2)=0$

$\d_\nu u_1A^\nu\Phi_1\Phi_3:\> (8/3-1/2)=d_QN_8$

$u_1A^\nu W_1^\nu\Phi_3:\> (8/3-3/4)=d_QN_8$

$u_3A^\nu W_1^\nu\Phi_1:\> (3/3-3/3)+(7/4-3/4)=d_QN_8$

$u_0Z^\nu\Phi_1\d_\nu\Phi_1:\> (7/2-2/1)+(7/2-2/2)=0$

$u_0\d_\nu Z^\nu\Phi_1^2:\> (7/2-2/2)=0$

$\d_\nu u_0 Z^\nu\Phi_1^2:\> (7/2-2/2)=d_QN_9$

$u_3A^\nu\Phi_1\d_\nu\Phi_1:\> (7/4-1/1)+(7/4-1/2)=0$

$u_3\d_\nu A^\nu\Phi_1^2:\> (7/4-1/2)=0$

$\d_\nu u_3A^\nu\Phi_1^2:\> (7/4-1/2)=d_QN_9$

$u_1A^\nu Z^\nu\Phi_1:\> (2/4-4/1)+(3/4-3/3)=d_QN_9$

$u_0W_1^\nu Z^\nu\Phi_1:\> (2/3-4/1)+(7/2-4/2)=0$

$\tilde u_3u_0u_1\Phi_1:\> (5/1-5/3)+(7/2-5/4)=0$

$\tilde u_1u_0u_3\Phi_1:\> (7/2-6/2)+(7/4-4/3)=0$

$\tilde u_1u_0u_1\Phi_3:\> (5/1-5/2)+(8/3-4/3)=0$
$$N_8=-ig^2\sin\Theta \Phi_3A_\nu(W_1^\nu\Phi_1+W_2^\nu\Phi_2)\delta(x-y) 
\eqno(A.8)$$
$$N_9=-ig^2(\tan\Theta-\sin 2\Theta)A_\nu Z^\nu(\Phi_1^2+\Phi_2^2)\delta(x-y)
\eqno(A.9)$$
{\it 7) Sector (1,1,1,1)}:

$u_1W_1^\nu\Phi_1\d_\nu\Phi_1:\> (10/2-9/1)+(10/2-9/2)=0$

$u_1\d_\nu W_1^\nu\Phi_1^2:\> (10/2-9/2)=0$

$\d_\nu u_1W_1^\nu\Phi_1^2:\> (10/2-9/2)=d_QN_{10}$

$u_1W_1^\nu W_{1\nu}\Phi_1:\> (10/2-9/3)=d_QN_{10}$

$u_1\Phi_1^3:\> (10/2-10/1)=d_QN_{11}$

$$N_{10}={i\over 4}g^2(W_{1\nu}W_1^\nu\Phi_1^2+W_{2\nu}W_2^\nu\Phi_2^2) 
\delta(x-y)\eqno(A.10)$$
$$N_{11}=-ig^2{m_H^2\over 16m_W^2}(\Phi_1^4+\Phi_2^4)\delta(x-y)\eqno(A.11)$$
{\it 8) Sector (1,1,2,2)}:

$u_1W_2^\nu\Phi_1\d_\nu\Phi_2:\> (1/1-1/2)+(4/1-2/2)+(8/2-3/2)+ 
(10/2-10/3)+(10/2-10/3)=0$

$u_1W_2^\nu\d_\nu \Phi_1\Phi_2:\> (1/1-1/1)+(4/1-2/1)+(8/2-3/1)+ 
(8/2-3/2)+(10/2-10/3)=0$

$u_1\d_\nu W_2^\nu\Phi_1\Phi_2:\> (8/2-3/2)+(10/2-10/3)=0$

$\d_\nu u_1W_2^\nu\Phi_1\Phi_2:\> (8/2-3/2)+(10/2-10/3)=0$

$u_1\Phi_1\Phi_2^2:\> (10/2-10/3)=d_QN_{12}$

$\tilde u_2u_1u_2\Phi_1:\> (5/3-4/4)+(6/3-6/1)+(8/1-5/1)+(10/2-10/3)=0$

$u_1W_{1\nu}W_2^\nu\Phi_2:\> (1/1-3/4)+(4/1-4/2)=0$

$u_1W_1^\nu\Phi_2\d_\nu\Phi_2:\> (8/2-2/3)+(8/2-2/4)=0$

$u_1\d_\nu W_1^\nu\Phi_2^2:\> (8/2-2/4)=0$

$\d_\nu u_1W_1^\nu\Phi_2^2:\> (8/2-2/4)=d_QN_{13}$

$u_2W_{1\nu}W_1^\nu\Phi_2:\> (1/2-3/4)+(4/2-4/2)+(10/4-9/3)=d_QN_{13}$
$$N_{12}=-ig^2{m_H^2\over 8m_W^2}\Phi_1^2\Phi_2^2\delta(x-y)\eqno(A.11)$$
$$N_{13}={i\over 4}g^2(W_{1\nu}W_1^\nu\Phi_2^2+W_{2\nu}W_2^\nu\Phi_1^2) 
\delta(x-y)\eqno(A.13)$$
{\it 9) Sector (2,2,3,3)}:

$u_2Z^\nu\Phi_2\d_\nu\Phi_3:\>
(3/1-2/4)+(8/4-2/1)+(10/4-11/1)+(10/4-11/2)=0$

$u_2Z^\nu\d_\nu\Phi_2\Phi_3:\> (3/1-2/3)+(8/4-2/2)+(8/4-2/1)+(10/4-11/2)=0$

$u_2\d_\nu Z^\nu\Phi_2\Phi_3:\> (8/4-2/1)+(10/4-11/2)=0$

$\d_\nu u_2Z^\nu\Phi_2\Phi_3:\> (8/4-2/1)+(10/4-11/2)=d_QN_{14}$

$u_3W_2^\nu\Phi_2\d_\nu\Phi_3:\> (3/2-2/4)+(7/3-3/1)+(7/3-3/2)+(11/2-10/3)=0$

$u_3W_2^\nu\d_\nu\Phi_2\Phi_3:\> (3/2-2/3)+(7/3-3/1)+(11/2-10/3)+(11/2-10/3)=0$

$u_3\d_\nu W_2^\nu\Phi_2\Phi_3:\> (7/3-3/1)+(11/2-10/3)=0$

$\d_\nu u_3W_2^\nu\Phi_2\Phi_3:\> (7/3-3/1)+(11/2-10/3)=d_QN_{14}$

$u_2W_2^\nu Z_\nu\Phi_3:\> (8/4-4/1)=d_QN_{14}$

$u_3W_2^\nu Z_\nu\Phi_2:\> (7/3-4/1)+(3/2-4/2)=d_QN_{14}$

$u_3Z^\nu\Phi_2\d_\nu\Phi_2:\> (7/3-2/2)+(7/3-2/1)=0$

$u_3\d_\nu Z^\nu\Phi_2^2:\> (7/3-2/1)=0$

$\d_\nu u_3Z^\nu\Phi_2^2:\> (7/3-2/1)=d_QN_{15}$

$u_2Z^\nu Z_\nu\Phi_2:\> (3/1-4/2)+(10/4-11/3)=d_QN_{15}$

$u_2W_2^\nu\Phi_3\d_\nu\Phi_3:\> (8/4-3/2)+(8/4-3/1)=0$

$u_2\d_\nu W_2^\nu\Phi_3^2:\> (8/4-3/1)=0$

$\d_\nu u_2W_2^\nu\Phi_3^2:\> (8/4-3/1)=d_QN_{16}$

$u_3W_2^\nu W_2^\nu\Phi_3:\> (11/2-10/3)=d_QN_{16}$

$u_2\Phi_2\Phi_3^2:\> (10/4-12/1)=d_QN_{17}$

$u_3\Phi_2^2\Phi_3:\> (11/2-10/3)=d_QN_{17}$

$\tilde u_2u_2u_3\Phi_3:\> (6/1-5/1)+(8/4-6/1)+(11/2-10/3)=0$

$\tilde u_3u_2u_3\Phi_2:\> (6/1-5/4)+(7/3-5/3)+(10/4-12/2)=0$
$$N_{14}=ig^2\sin\Theta\tan\Theta Z_\nu\Phi_3(W_1^\nu\Phi_1
+W_2^\nu\Phi_2)\delta(x-y)\eqno(A.13)$$
$$N_{15}=ig^2\Bigl(\cos\Theta-{1\over 2\cos\Theta}\Bigl)^2Z_\nu Z^\nu 
(\Phi_1^2+\Phi_2^2)\delta(x-y)\eqno(A.15)$$
$$N_{16}={i\over 4}g^2\Phi_3^2(W_{1\nu}W_1^\nu+W_{2\nu}W_2^\nu) 
\delta(x-y)\eqno(A.16)$$
$$N_{17}=-ig^2{m_H^2\over 8m_W^2}(\Phi_1^2+\Phi_2^2)\Phi_3^2 
\delta(x-y)\eqno(A.17)$$
{\it 10) Sector (3,3,3,3)}:

$u_3Z^\nu\Phi_3\d_\nu\Phi_3:\> (11/2-11/1)+(11/2-11/2)=0$

$u_3\d_\nu Z^\nu\Phi_3^2:\> (11/2-11/2)=0$

$\d_\nu u_3Z^\nu\Phi_3^2:\> (11/2-11/2)=d_QN_{18}$

$u_3Z_\nu Z^\nu\Phi_3:\> (11/2-11/3)=d_QN_{18}$

$u_3\Phi_3^3:\> (11/2-12/1)=d_QN_{19}$
$$N_{18}={ig^2\over 4\cos^2\Theta}Z_\nu Z^\nu\Phi_3^2\delta(x-y)\eqno(A.18)$$
$$N_{19}=-ig^2{m_H^2\over 16m_W^2}\Phi_3^4\delta(x-y).\eqno(A.19)$$
A further normalization term
$$N_{20}=ig^2\lambda\Phi_0^4\delta(x-y)\eqno(A.20)$$
is possible, but not necessary until now because $d_QN_{20}=0$. 
Similar to $b$ in (3.20.12),
the coupling constant $\lambda$ is arbitrary so far. But gauge
invariance at third order determines $b$ and $\lambda$. In this way we
can {\it derive} the Higgs potential. This will be discussed in part II. 
The other normalization terms $N_1-N_{19}$ agree precisely with
the four-legs vertices (i.e. the quartic part of the interaction Lagrangian)
of the standard model \cite{13}.

The normalization terms $N_2, N_4, N_{11}, N_{12}, N_{17}, N_{19}, N_{20}$
with four scalar fields are of a special kind which means that they do
not belong to a second order tree graph with two internal
derivatives as in footnote 11, i.e. there are no
corresponding non-local terms of second order;
but at {\it higher orders} such terms appear, for example
in forth order box diagrams with all derivatives on internal lines 
(see \cite{2}, p.341). 
\vskip0.5cm
{\bf Acknowledgements:} We thank Klaus Fredenhagen and Bert Schroer for 
stimulating discussions.


\end{document}